\RequirePackage{fix-cm}
\documentclass[twocolumn,epjc3]{svjour3}  
\smartqed  
\RequirePackage{graphicx}
\RequirePackage{mathptmx}      
\RequirePackage{amsmath}
\RequirePackage{amssymb}
\RequirePackage{cite}
\RequirePackage{slashed}
\RequirePackage{subfigure}
\RequirePackage{longtable}
\RequirePackage[colorlinks,citecolor=blue,urlcolor=blue,linkcolor=blue]{hyperref}
\newcommand{\eq}[1]{\begin{align}#1\end{align}}
\renewcommand\d{\partial}

\renewcommand{\u}[1]{\textrm{U}(#1)}
\newcommand{\su}[1]{\textrm{SU}(#1)}
\newcommand\T{\rule{0pt}{2.6ex}}       
\newcommand\B{\rule[-1.2ex]{0pt}{0pt}} 
\newcommand{\nn}{\nonumber} 

\newcommand{\ba}{\begin{array}}
\newcommand{\ea}{\end{array}}
\newcommand{\disp}{\displaystyle}
\newcommand{\Lag}{{\cal L}}
\newcommand{\orden}{\sim}
\newcommand{\dis}{\hspace{0.08cm}}
\newcommand{\spc}{\hspace{0.15cm}}
\newcommand{\vev}{{\em vev}}
\newcommand{\hc}{{\rm h.c.}}
\allowdisplaybreaks
\journalname{Eur. Phys. J. C}
\begin{document}

\title{A new and gauge-invariant Littlest Higgs model with T-parity
}


\author{Jos\'e Ignacio Illana\thanksref{e1,addr1}
        \and
        Jos\'e Mar\'ia P\'erez-Poyatos\thanksref{e2,addr1}
}

\thankstext{e1}{e-mail: jillana@ugr.es}
\thankstext{e2}{e-mail: jmppoyatos@ugr.es}


\institute{CAFPE and Departamento de F{\'\i}sica Te\'orica y del Cosmos,
           Universidad de Granada, E-18071 Granada, Spain \label{addr1}
}

\date{Received: date / Accepted: date}

\maketitle

\begin{abstract}
We inspect the Littlest Higgs model with T-parity, based on a global symmetry SU(5) spontaneously broken to SO(5), in order to elucidate the pathologies it presents due to the non trivial interplay between the gauge invariance associated to the heavy modes and the discrete T-parity symmetry. In particular, the usual Yukawa Lagrangian responsible for providing masses to the heavy `mirror' fermions is not gauge invariant. This is because it contains an SO(5) quintuplet of right-handed fermions that transforms non-linearly under SU(5), hence involving in general all SO(5) generators when a gauge transformation is performed and not only those associated to its gauge subgroup. Part of the solution to this problem consists of completing the right-handed fermion quintuplet with T-odd `mirror partners' and a gauge singlet, what has been previously suggested for other purposes. Furthermore, we find that the singlet must be T-even, the global symmetry group must be enlarged, an additional non-linear sigma field should be introduced to parametrize the spontaneous symmetry breaking and new extra fermionic degrees of freedom are required to give a mass to all fermions in an economic way while preserving gauge invariance. Finally, we derive the Coleman-Weinberg potential for the Goldstone fields using the background field method.
\end{abstract}


\section{Introduction}\label{Introduction}

Besides supersymmetry, Composite Higgs models \cite{panicoCompositeNambuGoldstoneHiggs2015} are one of the most elegant proposals to alleviate the fine-tuning in the Higgs mass afflicting the Standard Model (SM). In this family of models, the Higgs boson arises as the pseudo-Nambu-Goldstone boson (pNGB) of a spontaneously broken global symmetry. Within this class of models, one of the most popular frameworks is the \textit{Littlest Higgs model with T-parity} (LHT) \cite{arkani-hamedConstructingDimensions2001,arkani-hamedElectroweakSymmetryBreaking2001,ArkaniHamed:2002qy,chengTeVSymmetryLittle2003,chengLittleHierarchyLittle2004,lowParityLittlestHiggs2004,chengTopPartnersLittle2006} based on a global symmetry group SU(5) spontaneously broken to an SO(5) subgroup by the vacuum expectation value (\vev) of a symmetric tensor field $\Sigma$ at a scale $f \orden 1-10$ TeV. Using the Callan-Coleman-Wess-Zumino (CCWZ) formalism \cite{colemanStructurePhenomenologicalLagrangians1969,callanStructurePhenomenologicalLagrangians1969}, a non-linear field $\xi$ is introduced, built as the exponential of the 14 Goldstone bosons in the direction of the broken generators. Since the coset $\textrm{SU(5)/SO(5)}$ is symmetric, there is an inherited $Z_2$ automorphism in the algebra, allowing the definition of the T-parity discrete symmetry. This T-parity symmetry significantly relaxes direct and indirect constraints from electroweak precision data (EWPD) \cite{hubiszPhenomenologyLittlestHiggs2005,hubiszElectroweakPrecisionConstraints2006}: the SM particles are T-even and (most of) the new particles are T-odd and hence pair-produced. The Goldstone sector includes a physical T-even doublet, which is identified with the SM Higgs, and an extra T-odd physical triplet with zero \vev\ or otherwise T-parity would be broken. A subgroup $\left[\su{2}\times \u{1}\right]^2$ of the full global symmetry group SU(5) is gauged and gets broken spontaneously to the diagonal subgroup $\su{2}\times \u{1}$, leading to a set of T-odd massive vector bosons with masses of order $f$ after eating up the rest of Goldstone scalar fields. The Higgs mass, apart from being protected by the T-parity, is also protected by the so called \textit{collective symmetry breaking} mechanism \cite{hubiszPhenomenologyLittlestHiggs2005,schmaltzLittleHiggsReview2005}: the global symmetry is also broken explicitly by gauge and Yukawa interactions, 
but this breaking only occurs when two or more couplings are not vanishing at the same time. Otherwise there would still remain an unbroken global symmetry that is sufficient to ensure the Goldstone nature of the Higgs boson. When those couplings are not set to zero the model predicts a divergence in the Higgs mass sensitive to just the logarithm of the cut-off, hence solving the problems of quadratic divergences and fine-tuning of the SM.

In spite of its success to cope with the issues above, the LHT suffers pathologies in the fermionic sector (except for the third-generation quarks \cite{hubiszPhenomenologyLittlestHiggs2005,Han:2005ru}). The matter content, that includes extra fermions, breaks explicitly the gauge invariance of the model, as was already pointed out in \cite{Csaki:2008se,pappadopuloTparityItsProblems2011}. In most phenomenological studies  \cite{Belyaev:2006jh,Blanke:2006sb,blankeRareCPViolatingDecays2007,delAguila:2008zu,delAguila:2010nv,gotoTauMuonLepton2011,zhouFlavorChangingTop2013,yangLeptonFlavorViolating2017}, the left-handed SM and the so-called {\em mirror} fermions are introduced in incomplete multiplets of SU(5) whereas their right-handed counterparts come in multiplets of SO(5) that include an \su{2} doublet of  mirror fermions (sometimes completed with a SU(2) doublet of {\em mirror partners} and a {\em gauge singlet}). The mirror fermions acquire a vector-like mass by a Yukawa coupling to the non-linear sigma field $\xi$ and the rest (when they are not ignored) need extra mass terms. However, as we will show, this assumption breaks gauge invariance, due to the non-linear transformation of the SO(5) multiplet that mixes all the right-handed fermionic fields, being impossible to separate them to give different masses to its components. On the other hand, it is usually claimed that the T-parity of the singlet field can be chosen to be either odd \cite{chengLittleHierarchyLittle2004,lowParityLittlestHiggs2004,Reuter:2013iya} or even \cite{delaguilaLeptonFlavorChanging2017,delaguilaInverseSeesawNeutrino2019}, what gives rise to a very different phenomenology \cite{delaguilaInverseSeesawNeutrino2019}. But again a close look will reveal that only a T-even singlet is compatible with gauge invariance. In this work we propose an economic cure to these problems which consists in enlarging minimally the global group, introducing a new pattern for the spontaneous symmetry breaking (SSB) and adding extra fermionic degrees of freedom.

Once the new Lagrangian is built, we derive the Coleman-Weinberg potential for the scalar fields \cite{colemanRadiativeCorrectionsOrigin1973} following the background field method \cite{ballChiralGaugeTheory1989,dennerApplicationBackgroundFieldMethod1995,dittmaierDerivingNondecouplingEffects1995,buchallaCompleteOneLoopRenormalization2018,buchallaMasterFormulaOneloop2019}, that allows for the calculation of the divergent part of the potential in terms of the non-linear sigma fields of the theory.
The logarithmic divergences were already given by a master formula \cite{buchallaCompleteOneLoopRenormalization2018,buchallaMasterFormulaOneloop2019} that was derived rewriting the one-loop effective action in the Schwinger representation and applying a heat kernel expansion in the so-called proper time variable using dimensional regularization. However, we are interested in distinguishing between quadratic and logarithmic divergences so we will rather impose a cut-off in the proper time \cite{tomsQuadraticDivergencesQuantum2011}. We will reproduce the previous formula for the logarithmic part and find a new master formula for the quadratic divergences in the cut-off regularization scheme.

This paper is organized as follows. In section~\ref{review} we review the usual LHT to fix the notation. Section~\ref{pathologies} contains a detailed explanation of the pathologies of the model. Section~\ref{new proposal} is devoted to address the issues and construct the Lagrangian of a new and gauge invariant Littlest Higgs model with T-parity. In section~\ref{BFM} we introduce the background field method to evaluate the Coleman-Weinberg potential for the Goldstone fields. Finally, in the last section we present our conclusions and outlook.

\section{The Littlest Higgs model with T-parity in a nutshell}\label{review}

\subsection{Global symmetries \label{sec:su5toso5}}

In this section we review the usual LHT, following closely the notation of refs.~\cite{delaguilaLeptonFlavorChanging2017,delaguilaInverseSeesawNeutrino2019}. The model is based on the symmetric coset $\textrm{SU(5)/SO(5)}$ parametrized by the \vev\ of a symmetric tensor, 
\eq{
\Sigma_{0}=\left(\begin{array}{ccc}
0_{2\times 2} & 0 & {\bf{1}}_{2\times 2}\\
0 & 1 & 0\\
{\bf{1}}_{2\times2} & 0 & 0_{2\times2}
\end{array}\right),
\label{Sigma0}
}
leaving $24-10=14$ unbroken generators. This spontaneous breaking direction fixes the embedding of SO(5) in SU(5) with the fundamental representation of the latter reduced to
the defining (real) representation of the former. The unbroken generators preserve the vacuum verifying the relation
\eq{\label{unbroken}
T^a\Sigma_0+\Sigma_0T^{a T}=0.
}
The expression above suggests the definition of an automorphism in the Lie algebra under which the unbroken generators transform as \cite{chengLittleHierarchyLittle2004}
\eq{
T^a \xrightarrow{\textrm{aut}} -\Sigma_0T^{a T}\Sigma_0=T^a,
}
where the last equality follows from eq.~(\ref{unbroken}). The set of broken generators will be orthogonal to the unbroken ones if their eigenvalue under the automorphism is the opposite,
\eq{
X^a \xrightarrow{\textrm{aut}} -\Sigma_0X^{a T}\Sigma_0=-X^a.
}
This characterizes the broken generators as the set that verifies
\eq{\label{broken}
X^a\Sigma_0-\Sigma_0X^{a T}=0.
}
Broken and unbroken generators verify the following schematic commutation relations,
\eq{
\left[T,T\right]\orden T,\quad
\left[T,X\right]\orden X,\quad
\left[X,X\right]\orden T,
}
that can be derived from the automorphism. One usually takes an orthogonal basis of generators. The broken generators expand the Goldstone matrix $\Pi=\pi^a X^a$ and allows the introduction of a non-linear field $\xi$ that transforms under the global symmetry group,
\eq{\label{xitransformation}
\xi= e^{i\Pi/f},\quad \xi\xrightarrow{G} V\xi U^{\dagger}
}
where $f$ is the scale of new physics, $V$ is an SU(5) transformation and $U=U\left(V,\Pi\right)$ is the compensating SO(5) non-linear transformation, that depends on $V$ and $\Pi$. According to the CCWZ formalism, the transformation of $\xi$ is such that it keeps the exponential form. 

Let us derive a relevant property that $U$ satisfies. The characterization of the broken generators (\ref{broken}) leads to
\eq{\label{xi}
\xi=\Sigma_0\xi^T\Sigma_0
}
and applying the transformation given by the CCWZ formalism to this expression one finds
\eq{\label{Udef}
V\xi U^{\dagger}
=\Sigma_0(V\xi U^{\dagger})^T\Sigma_0
=U\xi\Sigma_0V^T\Sigma_0,
}
where we have used eq.~(\ref{xi}), the hermiticity of the generators and the defining property of the unbroken ones in eq.~(\ref{unbroken}) that implies $U\Sigma_0=\Sigma_0 U^{*}$. Equation~(\ref{Udef}) can be interpreted as a definition for the non-linear transformation $U$ and it is also a consequence of the spontaneous breaking of SU(5) to SO(5). 

This formalism allows to define a tensor field $\Sigma$ that transforms linearly under the global symmetry group
\eq{
\Sigma=\xi\Sigma_0\xi^{T}=\xi^2\Sigma_0,\quad \Sigma\xrightarrow{G} V\Sigma V^{T},
}
where we have used eq.~(\ref{unbroken}) to commute $\xi$ with $\Sigma_0$.

\subsection{Gauge group}

The group SU(5) contains two copies of \su{2}$\times$\u{1}. In order to implement the collective symmetry breaking mechanism in the gauge sector, one gauges the subgroup $G_g=\left[\su{2}\times\u{1}\right]_1\times \left[\su{2}\times \u{1}\right]_2$ expanded by the Hermitian and traceless generators
\eq{
Q_{1}^{a}&=\frac{1}{2}\left(\begin{array}{ccc}
\sigma^{a} & 0 & 0\\
0 & 0 & 0\\
0 & 0 & 0_{2\times2}
\end{array}\right) 
,\quad Y_1= \frac{1}{10}\textrm{diag}\left(3,3,-2,-2,-2\right),
\label{generators1}
\\
Q_{2}^{a}&=\frac{1}{2}\left(\begin{array}{ccc}
0_{2\times 2} & 0 & 0\\
0 & 0 & 0\\
0 & 0 & -\sigma^{a*}
\end{array}\right)
,\quad Y_2= \frac{1}{10}\textrm{diag}\left(2,2,2,-3,-3\right),
\label{generators2}
}
with $\sigma^a$ the three Pauli matrices. The normalization of the gauge generators is $\textrm{tr}\left(Q^a_j Q^b_k\right)=\frac{1}{2}\delta^{ab}\delta_{jk}$ and $\textrm{tr}\left(Y_j Y_k\right)=\frac{1}{10}\delta_{jk}+\frac{1}{5}$ and the rest of the traces vanish. A useful property that the gauge generators verify is
\eq{\label{propiedad}
Q^a_1=-\Sigma_0 Q^{a T}_2\Sigma_0,\quad Y_1=-\Sigma_0 Y^{T}_2\Sigma_0,
}
which relates the generators of both gauged subgroups. The \vev\ along the direction of $\Sigma_0$ also breaks spontaneously the gauge group down to the diagonal subgroup $\su{2}_L\times\u{1}_Y$ identified as the SM gauge group, generated by the combinations $\left\{Q_1^a+Q_2^a,Y_1+Y_2\right\}\subset \left\{T^a\right\}$, while the broken combinations $\left\{Q_1^a-Q_2^a,Y_1-Y_2\right\}\subset \left\{X^a\right\}$ expand the Goldstone matrix
\eq{
\scriptsize
\Pi=\left(\begin{array}{ccccc}-\disp\frac{\omega^0}{2}-\frac{\eta}{\sqrt{20}} & -\disp\frac{\omega^+}{\sqrt{2}} & -i\disp\frac{\pi^+}{\sqrt{2}} & -i\Phi^{++} & -i\disp\frac{\Phi^+}{\sqrt{2}} \\
-\disp\frac{\omega^-}{\sqrt{2}} & \disp\frac{\omega^0}{2}-\frac{\eta}{\sqrt{20}} & \disp\frac{v+h+i\pi^0}{2} & -i\disp\frac{\Phi^+}{\sqrt{2}} & \disp\frac{-i\Phi^0+\Phi^P}{\sqrt{2}} \\
i\disp\frac{\pi^-}{\sqrt{2}} & \disp\frac{v+h-i\pi^0}{2} & \sqrt{\disp\frac{4}{5}}\eta & -i\disp\frac{\pi^+}{\sqrt{2}} &  \disp\frac{v+h+i\pi^0}{2} \\
i\Phi^{--} & i\disp\frac{\Phi^-}{\sqrt{2}} & i\disp\frac{\pi^-}{\sqrt{2}} & -\disp\frac{\omega^0}{2}-\frac{\eta}{\sqrt{20}} & -\disp\frac{\omega^-}{\sqrt{2}} \\
i\disp\frac{\Phi^-}{\sqrt{2}} & \disp\frac{i\Phi^0+\Phi^P}{\sqrt{2}} &  \disp\frac{v+h-i\pi^0}{2} & -\disp\frac{\omega^+}{\sqrt{2}} & \disp\frac{\omega^0}{2}-\frac{\eta}{\sqrt{20}}
\end{array}\right),
}
where $v$ is the Higgs \vev. Under the SM gauge group the Goldstone matrix decomposes as
\eq{
\Pi: 1_0\oplus 3_0\oplus 2_{1/2}\oplus 3_1,
}
including a complex symmetric $\su{2}$ triplet and its hermitian conjugate
\eq{
\small
\Phi=\left(\begin{array}{cc}
-i\Phi^{++} & -i\disp\frac{\Phi^{+}}{\sqrt{2}}\\
-i\disp\frac{\Phi^{+}}{\sqrt{2}} & \disp\frac{-i\Phi^{0}+\Phi^{P}}{\sqrt{2}}
\end{array}\right),\quad \Phi^{\dagger}=\left(\begin{array}{cc}
i\Phi^{--} & i\disp\frac{\Phi^{-}}{\sqrt{2}}\\
i\disp\frac{\Phi^{-}}{\sqrt{2}} & \disp\frac{i\Phi^{0}+\Phi^{P}}{\sqrt{2}}
\end{array}\right),
}
the SM Higgs doublet
\eq{
\label{Hdoublet}
H=\left(\begin{array}{c}
i\pi^{+}\\
\disp\frac{v+h+i\pi^{0}}{\sqrt{2}}
\end{array}\right),
}
plus a $\su{2}$ triplet 
\eq{\omega=\left(\begin{array}{cc}
-\disp\frac{\omega^0}{2} & -\disp\frac{\omega^+}{\sqrt{2}}\\
-\disp\frac{\omega^-}{\sqrt{2}} & \disp\frac{\omega^0}{2}
\end{array}\right)
} 
and a singlet ($\eta$). The latter two will become the longitudinal modes of the heavy gauge fields. 

\subsection{Lagrangian}

\subsubsection{Gauge sector}

In the construction of the Lagrangian we take into account the action of the discrete T-parity symmetry which is introduced to keep the SM gauge bosons T-even and light while the new ones are T-odd and heavy. The action of T-parity consists in an interchange of the two gauge groups
\eq{
G_1\overset{\rm T}{\longleftrightarrow} G_2,
}
where $G_1=\left[\su{2}\times\u{1}\right]_1$ and $G_2=\left[\su{2}\times\u{1}\right]_2$. This requires that the coupling constants of both copies must be the same $g_1=g_2=\sqrt{2}g$, $g'_1=g'_2=\sqrt{2}g'$, with the first set of couplings referring to \su{2} and the second to \u{1}. In this way, the T-parity affects the collective symmetry breaking in the gauge sector, since being the couplings equal for both subgroups they are different from zero at the same time. The gauge Lagrangian takes the usual form
\eq{
\Lag_{G}=\sum_{j=1}^2\left[-\frac{1}{2}\textrm{tr}\left(\widetilde{W}_{j\mu\nu}\widetilde{W}_j^{\mu\nu}\right)-\frac{1}{4}B_{j\mu\nu}B_j^{\mu\nu}\right],
\label{lagG}
}
in terms of fields and field strength tensors,
\eq{
\widetilde{W}_{j\mu}&=W^a_{j\mu}Q^a_j,\quad \\ \widetilde{W}_{j\mu\nu}&=\partial_{\mu}\widetilde{W}_{j\nu}-\partial_{\nu}\widetilde{W}_{j\mu}-i\sqrt{2}g\left[\widetilde{W}_{j\mu},\widetilde{W}_{j\nu}\right],\quad \\
B_{j\mu\nu}&=\partial_{\mu}B_{j\nu}-\partial_{\nu}B_{j\mu},
}
where in the first expression the index $j$ is fixed. Before the electroweak SSB, the SM gauge bosons come from the T-even combinations
\eq{
W^{\pm}&=\frac{1}{2}\left[\left(W^1_1+W^1_2\right)\mp i\left(W^2_1+W^2_2\right)\right],\quad \\
W^3&=\frac{W^3_1+W^3_2}{\sqrt{2}},\quad
B=\frac{B_1+B_2}{\sqrt{2}},
\label{gaugeTeven}
}
while the remaining T-odd combinations will define the heavy fields
\eq{
W_H^{\pm}&=\frac{1}{2}\left[\left(W^1_1-W^1_2\right)\mp i\left(W^2_1-W^2_2\right)\right],\quad \\
W_H^3&=\frac{W^3_1-W^3_2}{\sqrt{2}},\quad
B_H=\frac{B_1-B_2}{\sqrt{2}}.
\label{gaugeTodd}
}

\subsubsection{Scalar sector}

In order to assign a T-even parity to the SM Higgs boson and T-odd parities to the rest of the scalar fields, one defines
\eq{
\Pi\xrightarrow{\textrm{T}} -\Omega \Pi \Omega, \quad \Omega=\textrm{diag}\left(-1,-1,1,-1,-1\right).
}
It is important to remark that $\Omega$ is an element of the center of the gauge subgroup and consequently only commutes with the gauge generators. This fact will be crucial in the following. The T-parity transformation of the Goldstone matrix implies
\eq{
\xi\xrightarrow{\textrm{T}}\Omega\xi^{\dagger}\Omega,\quad \Sigma\xrightarrow{\textrm{T}}\widetilde{\Sigma}\equiv\Omega\Sigma_0\Sigma^{\dagger}\Sigma_0\Omega.
} 
With these ingredients one builds the scalar Lagrangian which is gauge and T-parity invariant using eq.~(\ref{propiedad}), 
\eq{
\Lag_{S}=\frac{f^2}{8}\textrm{tr}\left[\left(D^{\mu}\Sigma\right)^{\dagger}D_{\mu}\Sigma\right],
\label{lagS}
}
where the covariant derivative is defined as
\eq{
D_{\mu}\Sigma&=\partial_{\mu}\Sigma-\sqrt{2}i\sum_{j=1}^2\big[g W^a_{j\mu}\left(Q^a_j\Sigma+\Sigma Q^{a T}_j\right) \nn\\
&-g'B_{j\mu}\left(Y_j\Sigma+\Sigma Y_j^T\right)\big].
\label{scalarderivative}
}

\subsubsection{Fermion sector}

Implementing T-parity and giving masses to all the fermions is less straightforward. In fact, this is the main source of the trouble we will address and try to solve. Here we will focus on the leptonic sector but the same construction applies to quarks, except for the top quark, that has extra couplings and additional partners.

First of all, in the usual procedure \cite{chengLittleHierarchyLittle2004,lowParityLittlestHiggs2004} one introduces two left-handed SU(5) quintuplets in the anti-fundamental and fundamental representations, respectively, 
\eq{
\Psi_1=\left(\begin{array}{c}
-i\sigma^{2}l_{1L}\\
0\\
0_{2}
\end{array}\right),\quad \Psi_2=\left(\begin{array}{c}
0_2\\
0\\
-i\sigma^{2}l_{2L}
\end{array}\right).
}
The explicit form of these multiplets breaks explicitly the global SU(5) symmetry, because they are incomplete, but the gauge subgroup is preserved. In particular, under a gauge transformation,
\eq{
\Psi_1\xrightarrow{G_g} V_g^{*}\Psi_1,\quad \Psi_2\xrightarrow{G_g} V_g\Psi_2.
}
For a T-parity transformation it is common to contemplate two options \cite{chengLittleHierarchyLittle2004,lowParityLittlestHiggs2004,delaguilaLeptonFlavorChanging2017,delaguilaInverseSeesawNeutrino2019}:
\eq{
&a)\quad\Psi_1\xrightarrow{\textrm{T}}\Omega\Sigma_0\Psi_2,\label{Tevenfermions}\\
&b)\quad\Psi_1\xrightarrow{\textrm{T}}-\Sigma_0\Psi_2.
}
Then one can define T-even and T-odd combinations given respectively by
\eq{
a)\quad&\Psi_{+}=\frac{\Psi_1+\Omega\Sigma_0\Psi_2}{\sqrt{2}},\quad \Psi_{\mathunderscore}=\frac{\Psi_1-\Omega\Sigma_0\Psi_2}{\sqrt{2}},
\label{Tevenchi}\\
b)\quad&\Psi_{+}=\frac{\Psi_1-\Sigma_0\Psi_2}{\sqrt{2}},\quad \Psi_{\mathunderscore}=\frac{\Psi_1+\Sigma_0\Psi_2}{\sqrt{2}}.
}
The T-odd combination $l_{HL}=(l_{1L}+l_{2L})/\sqrt{2}$ needs to be paired with a right-handed doublet $l_{HR}$ so that `mirror' leptons $l_H$ may get a vector-like mass.
To that end, a right-handed SO(5) quintuplet is introduced,
\eq{
\Psi_{R}=\left(\begin{array}{c}
-i\sigma^{2}(\widetilde{l}^{c}_{\mathunderscore})_{R}\\
i\chi_{R}\\
-i\sigma^{2}l_{HR}
\end{array}\right).
\label{psiR}
}
We will denote with a subscript $\pm$ the T-parity assignment to be defined below.
The T-odd doublet $(\widetilde{l}^c_{\mathunderscore})_R$ describes the `mirror partner' leptons and $\chi_R$ is a gauge singlet, that in principle can be taken either $(\chi_+)_R$ or $(\chi_-)_R$.  Some authors \cite{gotoTauMuonLepton2011,zhouFlavorChangingTop2013,yangLeptonFlavorViolating2017,blankeRareCPViolatingDecays2007} leave this quintuplet incomplete, including in it just the doublet $l_{HR}$.  The transformation under the gauged subgroup reads
\eq{\label{PsiRtransformation}
\Psi_R \xrightarrow{G_g} U_g\Psi_R,
}
where $U_g$ is an SO(5) non linear transformation that verifies eq.~(\ref{Udef}) for a given $V_g$. Under T-parity, there are two possible realizations:
\eq{
a)\label{TevenPsiRtransformation}\quad&\Psi_R\xrightarrow{\textrm{T}}\Omega\Psi_R,\\
b)\quad&\Psi_R\xrightarrow{\textrm{T}}-\Psi_R.
}
The first one differs from the second in that $\chi_R$ is T-even; the rest of the fields are all T-odd. With this in mind, one can correspondingly construct two versions of a Yukawa Lagrangian,
\eq{
\Lag_{Y_H}^{(a)}&=-\kappa f\left(\overline{\Psi}_2\xi+\overline{\Psi}_1\Sigma_0\xi^{\dagger}\right)\Psi_R+\hc,\label{lagkappaa}\\
\Lag_{Y_H}^{(b)}&=-\kappa f\left(\overline{\Psi}_2\xi+\overline{\Psi}_1\Sigma_0\Omega\xi^{\dagger}\Omega\right)\Psi_R+\hc\label{lagkappab}
}
tailored to provide the mirror leptons with a mass of order $\kappa f$. 

In order to give a mass of order $\lambda v$ to the SM leptons after the electroweak SSB, the following Yukawa Lagrangian has been proposed \cite{hubiszPhenomenologyLittlestHiggs2005,
hanNeutrinoMassesLeptonnumber2005,hubiszFlavorLittleHiggs2005,hubiszElectroweakPrecisionConstraints2006,chenHiggsBosonProduction2006,Goto:2008fj},
\eq{
\Lag_{Y}&=i\frac{\lambda f}{4}\epsilon_{xyz}\epsilon_{rs}\left[(\overline{\Psi_2^{X^{*}}})_x\Sigma_{ry}\Sigma_{sz}+(\overline{\Psi_1^{X}}\Sigma_0\Omega)_x\widetilde\Sigma_{ry}\widetilde\Sigma_{sz}\right]\ell_R\nn\\&+\hc,
\label{laglambda}
}
where $\left\{x,y,z\right\}=3,4,5$ and $\left\{r,s\right\}=1,2$. Here the left-handed fermions are embedded in incomplete SU(5) quintuplets,
\eq{\label{PsiX}
\Psi_1^{X}=\left(\begin{array}{c}
X l_{1L}\\
0\\
0_{2}
\end{array}\right),\quad \Psi_2^{X^{*}}=\left(\begin{array}{c}
0_2\\
0\\
X^{*} l_{2L}
\end{array}\right),
}
transforming under the gauge group as
\eq{\label{trans1}
\Psi_1^{X} \xrightarrow{G_g} V_g \Psi_1^{X},\quad 
\Psi_2^{X^{*}} \xrightarrow{G_g} V_g^{*} \Psi_2^{X^{*}}
}
and under T-parity as
\eq{
\Psi_1^{X} \xrightarrow{\textrm{T}} 
\Omega\Sigma_0 \Psi_2^{X^{*}} = -\Sigma_0\Psi_2^{X^{*}},
}
where the last equality follows from the field content of these quintuplets. Despite the $\ell_R$ is an SU(5) singlet and all indices are contracted, this Lagrangian is not invariant under the global SU(5) symmetry, broken due to $\Sigma_0$ and the incomplete multiplets. Nevertheless we just need that the gauge symmetry $G_g$ is preserved, and this requires the introduction of $\Psi_1^X\,(\Psi_2^{X^*})$ transforming {\it opposite} to $\Psi_1\,(\Psi_2)$ so that the SM charged leptons, with left-handed components in the T-even doublet $l_{L}=(l_{1L}-l_{2L})/\sqrt{2}$, get a mass.\footnote{The neutrinos are massless throughout this work.} On the other hand, $\ell_R$ inherits no hypercharge from the global symmetry group. To give it the proper hypercharge ($Y=-1$) one has to enlarge SU(5) with two extra factors $\u{1}''_1$ and $\u{1}''_2$ hence preserving gauge invariance  \cite{gotoTauMuonLepton2011},
\eq{
\textrm{SU(5)}\times \u{1}_1''\times \u{1}_2''.
\label{extsu5}
}
Then for any field the actual hypercharges under the gauged $\u{1}_1$ and $\u{1}_2$ will be the sum of those under the $\u{1}'_1$ and $\u{1}'_2$ present in SU(5) plus the extra ones. The auxiliary field $X$ and its complex conjugate $X^{*}$ are introduced in eq.~(\ref{PsiX}) in order to reverse the U(1) charges of the left-handed components and at the same time compensate for the hypercharge assignment of the right-handed leptons. From the hypercharges of the $\Sigma$ components in table~\ref{table:hypercharge1} (note that all relevant products have the same values) and the requirements above one derives the charge assignments for the fermion fields of table~\ref{table:hypercharge2}. A particular realization of the scalar $X$ can be constructed with the fields already present in the model \cite{hubiszPhenomenologyLittlestHiggs2005,chenHiggsBosonProduction2006,gotoTauMuonLepton2011}. The element $\Sigma_{33}$ has hypercharges $\left(Y_1,Y_2\right)=(-\frac{2}{5},\frac{2}{5})$, and it is a \su{2} singlet, so we can identify $X=\Sigma_{33}^{-\frac{1}{4}}$ and $X^{*}=(\Sigma^{\dag}_{33})^{-\frac{1}{4}}$, which also verifies the right transformation under T-parity $X\xrightarrow{\textrm{T}}X^{*}$.\footnote{In view of the charge assignments in table \ref{table:hypercharge2}, it is clear that $X^{*}$ is not the complex conjugate of $X$, since they do not have opposite hypercharges under the \u{1} factors. However only gauged hypercharges matter and they are actually opposite, so we prefer to keep this notation. Indeed, the particular realization for these fields shows that one is the Hermitian conjugate of the other.}

\begin{table}
\caption{Hypercharges under $\u{1}'_1 \times \u{1}'_2\subset SU(5)$ of the $\Sigma$ components in eq.~(\ref{laglambda}).}\label{table:hypercharge1}
\centering
\begin{tabular}{|c|cc|}
\hline
 & \quad $Y'_1$ \quad & \quad $Y'_2$  \quad \T\B\\ 
\hline
$\Sigma_{13}$ & $\frac{1}{10}$ & $\frac{2}{5}$  \T\B\\ 
$\Sigma_{14}$ & $\frac{1}{10}$ & $-\frac{1}{10}$ \T\B\\ 
$\Sigma_{15}$ & $\frac{1}{10}$ & $-\frac{1}{10}$  \T\B\\
\hline
$\Sigma_{23}$ & $\frac{1}{10}$ & $\frac{2}{5}$   \T\B\\
$\Sigma_{24}$ & $\frac{1}{10}$ & $-\frac{1}{10}$ \T\B\\
$\Sigma_{25}$ & $\frac{1}{10}$ & $-\frac{1}{10}$ \T\B \\
\hline
$\Sigma_{13}\Sigma_{24}$ & $\frac{1}{5}$ & $\frac{3}{10}$  \T\B\\
$\Sigma_{13}\Sigma_{25}$ & $\frac{1}{5}$ & $\frac{3}{10}$  \T\B\\
$\Sigma_{23}\Sigma_{14}$ & $\frac{1}{5}$ & $\frac{3}{10}$  \T\B\\
$\Sigma_{23}\Sigma_{15}$ & $\frac{1}{5}$ & $\frac{3}{10}$  \T\B\\
\hline
\end{tabular}
\end{table}

\begin{table}
\caption{Hypercharge assignments under $\u{1}'_1 \times \u{1}'_2 \times \u{1}''_1 \times \u{1}''_2$, where the single-prime abelian groups are inside SU(5) and the others are extra factors. The hypercharges $Y = Y_1 + Y_2$ under the gauge group $\u{1}_1 \times \u{1}_2$ come from  $Y_j = Y'_j + Y''_j,\ j = 1,2$.}\label{table:hypercharge2}
\centering
\begin{tabular}{|c|cccc|cc|}
\hline
 & \quad $Y'_1$ \quad & \quad $Y'_2$ \quad & \quad $Y''_1$ \quad & \quad $Y''_2$ \quad & \quad $Y_1$ \quad & \quad $Y_2$ \T\B\\ 
\hline
$\ell_R$ & $0$ & $0$ & $-\frac{1}{2}$ &
$-\frac{1}{2}$ & $-\frac{1}{2}$ &
$-\frac{1}{2}$ \T\B\\ 
\hline
$ l_{2L}$ & $-\frac{1}{5}$ & $-\frac{3}{10}$ & $0$ & $0$ & $-\frac{1}{5}$ & $-\frac{3}{10}$  \T\B\\
$ l_{1L}$ & $-\frac{3}{10}$ & $-\frac{1}{5}$ & $0$ & $0$ & $-\frac{3}{10}$ & $-\frac{1}{5}$  \T\B\\
$ X^{*}l_{2L}$ & $\frac{1}{5}$ & $\frac{3}{10}$ & $-\frac{1}{2}$ & $-\frac{1}{2}$ & $-\frac{3}{10}$ & $-\frac{1}{5}$  \T\B\\
$ X l_{1L}$ & $\frac{3}{10}$ & $\frac{1}{5}$ & $-\frac{1}{2}$ & $-\frac{1}{2}$ & $-\frac{1}{5}$ & $-\frac{3}{10}$ \T\B \\
\hline
$X^*$ & $\frac{2}{5}$ & $\frac{3}{5}$ & $-\frac{1}{2}$ &
$-\frac{1}{2}$ & $-\frac{1}{10}$ & $\frac{1}{10}$ \T\B\\ 
$X$ & $\frac{3}{5}$ & $\frac{2}{5}$ & $-\frac{1}{2}$ & $-\frac{1}{2}$ & $\frac{1}{10}$ & $-\frac{1}{10}$\T\B\\ 
\hline
\end{tabular}
\end{table}

The mirror partner leptons $(\widetilde{l}^c_{\mathunderscore})$ and the gauge singlet ($\chi_+$ or $\chi_-$) must be heavy. It is customary \cite{chengLittleHierarchyLittle2004,lowParityLittlestHiggs2004} to give them a large vector-like mass introducing their left-handed components $(\widetilde{l}^c_{\mathunderscore})_L$ and $(\chi_{\pm})_L$ in incomplete SO(5) multiplets,
\eq{
\Psi_L = \left(\begin{array}{c}
(\widetilde{l}^c_-)_L\\
0\\
0_{2}
\end{array}\right), \quad
\Psi_L^\chi = \left(\begin{array}{c}
0\\
(\chi_\pm)_L\\
0_{2}
\end{array}\right).
\label{PsiL}
}
Their direct mass terms are assumed to be a {\it soft} breaking of the SO(5) global symmetry and have the form
\eq{\label{masswidetildechi}
\Lag_{M,M_{\chi}}=-M\overline{(\widetilde{l}^c_{\mathunderscore})}_L(\widetilde{l}^c_{\mathunderscore})_R-M_{\chi}\overline{(\chi_{\pm})}_L(\chi_{\pm})_R+\hc
}

Finally, the CCWZ formalism provides us with the kinetic terms and the gauge interactions of all fermions,
\eq{
\Lag_F = \Lag_{F_L} + \Lag_{F_R} + (\Psi_R \to \Psi_L) + (\Psi_R \to \Psi_L^\chi)
}
where
\eq{
\Lag_{F_L} =
i\overline{\Psi}_1\gamma^{\mu}D^{*}_{\mu}\Psi_1+i\overline{\Psi}_2\gamma^{\mu}D_{\mu}\Psi_2
\label{LagFL}
}
and depending on the T-parity implementation,
\eq{
\Lag_{F_R}^{(a)} &=
i\overline{\Psi}_R\gamma^{\mu}\left[\partial_{\mu}+\frac{1}{2}\xi^{\dag}\left(D_{\mu}\xi\right)+\frac{1}{2}\xi\Sigma_0D^{*}_{\mu}\left(\Sigma_0\xi^{\dag}\right)\right]\Psi_R,
\label{LagFR}
\\
\Lag_{F_R}^{(b)} &=
i\overline{\Psi}_R\gamma^{\mu}\big[\partial_{\mu}+\frac{1}{2}\xi^{\dag}\left(D_{\mu}\xi\right)\nn\\&+\frac{1}{2}\Omega\xi\Sigma_0D^{*}_{\mu}\left(\Sigma_0\xi^{\dag}\right)\Omega\big]\Psi_R,
}
with the covariant derivative defined as
\eq{
D_{\mu}&=\partial_{\mu}-\sqrt{2}i g \left(W^a_{1\mu} Q^a_1+W^a_{2\mu} Q^a_2\right) \nn\\
&+\sqrt{2}i g' \left(B_{1\mu}Y_1+B_{2\mu}Y_2\right).
\label{covderfer}
}
And for SM right-handed leptons, 
\eq{
\Lag_{F'}&=i\overline{\ell}_R\left[\partial_{\mu}-\sqrt{2}i g'\left(-\frac{1}{2}B_{1\mu}-\frac{1}{2}B_{2\mu}\right)\right]\ell_{R}
\nn\\ &=i\overline{\ell}_R\left(\partial_{\mu}+i g'B_{\mu}\right)\ell_{R}.
\label{LagFprime}
}


\section{Non gauge invariance of the original model}\label{pathologies}

This section is devoted to one of the main points of this work: the incompatibility between gauge invariance associated to the heavy modes and the usual T-parity implementation in which the gauge singlet $\chi$ is assumed T-odd. We will also show that in any case the content of the fermion multiplets transforming non linearly under SO(5) cannot be arbitrary in order to preserve the gauge invariance and T-parity at the same time.

Let us consider a generic global SU(5) transformation $V$ and find the corresponding non-linear transformation $U$ of SO(5), related by the spontaneous breaking of the first group down to the second. Both transformations can be parametrized by
\eq{
V=e^{i(\alpha^a X^a+\beta^b T^b)}, \quad
U=e^{i\sigma^b T^b},
}
where $U$ depends only on the SO(5) unbroken generators $T^b$ and $V$ depends on all the generators of SU(5). The fact that $U$ is a function of $V$ and the Goldstone fields $\Pi$ is encoded in the parameters $\sigma^b=\sigma^{b}(\alpha^a,\beta^b,\Pi)$. We can derive these parameters by taking infinitesimal transformations and using eq.~(\ref{Udef}):
\eq{
\left(1+i\alpha^a X^a+i\beta^b T^b\right)\xi&\left(1-i\sigma^b T^b\right) \nn\\
=\left(1+i\sigma^bT^b\right)&\xi\Sigma_0\left(1+i\alpha^a X^{a T}+i\beta^b T^{b T}\right)\Sigma_0\nn\\
\Rightarrow
\sigma^b \{T^b,\xi\}&=\beta^b \{T^b,\xi\}+\alpha^a\left[X^a,\xi\right],
\label{sigmaequation}
}
where we have used eqs.~(\ref{unbroken}) and (\ref{broken}) to eliminate $\Sigma_0$. As $\xi$ is a power expansion in the inverse of the high energy scale $f$, 
\eq{
\xi = e^{i\Pi/f} = \sum_{n=0}^\infty \frac{i^n}{n!}\frac{\Pi^n}{f^n},
}
one can assume the following ansatz for the parameters $\sigma^b$,
\eq{
\sigma^b=\sum_{n=0}^{\infty}\frac{\sigma^b_n}{f^n}.
}
At zero-th order in $1/f$, $\xi= 1$ and we find $\sigma^b_0=\beta^b$, which gives the linear part of the transformation. Particularizing for a gauge transformation, if higher order corrections were neglected, $U_g$ would only depend on the SM gauge generators and then the Lagrangians of eqs.~(\ref{lagkappaa}) and (\ref{lagkappab}) would be both gauge invariant, with or without completing the SO(5) quintuplet $\Psi_R$ with the gauge singlet and the partner leptons, because then the SM gauge generators would not mix the upper and lower components of the quintuplet. However this simplification cannot be done. To obtain the non-linear $\Pi$-dependent effects of the transformation, one must solve eq.~(\ref{sigmaequation}) for higher orders, 
\eq{
\sum_{n,m=0}^{\infty}\sigma_m^b\frac{1}{f^{n+m}}\frac{i^n}{n!}\{T^b,\Pi^n\}
&=\beta^b\sum_{n=0}^{\infty}\frac{1}{f^n}\frac{i^n}{n!}\{T^b,\Pi^n\} \nn\\&+
\alpha^a\sum_{n=0}^{\infty}\frac{1}{f^n}\frac{i^n}{n!}\left[X^a,\Pi^n\right].
}
Renaming the indices of the term in the l.h.s. we have
\eq{
\sum_{n,m=0}^{\infty}\sigma_m^b\frac{1}{f^{n+m}}&\frac{i^n}{n!}\{T^b,\Pi^n\}
\nn\\&=\sum_{n=0}^\infty\sum_{m=0}^n\sigma_m^b\frac{1}{f^n}\frac{i^{n-m}}{(n-m)!}\{T^b,\Pi^{n-m}\}
}
and then
\eq{
\sum_{m=0}^n\sigma_m^b\frac{i^{n-m}}{(n-m)!}\{T^b,\Pi^{n-m}\}
&=\beta^b\frac{i^n}{n!}\{T^b,\Pi^n\} \nn\\&+
\alpha^a\frac{i^n}{n!}\left[X^a,\Pi^n\right]
,\quad n\ge0,
}
that can be rewritten using that $\sigma_0^b=\beta^b$,
\eq{\label{sigmaequationexpanded}
\sum_{m=1}^{n}\sigma^b_m\frac{i^{n-m}}{\left(n-m\right)!}\{T^b,\Pi^{n-m}\}=\alpha^a \frac{i^n}{n!}\left[X^a,\Pi^n\right]
,\quad n\ge1,
}
where we see that the non linear part of the infinitesimal $U$ transformation only depends on the coefficients that go with the broken generators. With this expression we can obtain directly the coefficient $\sigma_1^b$ using that the basis of generators is orthogonal,
\eq{
2\sigma^b_1=i\alpha^a\textrm{tr}\left(\left[X^a,\Pi\right]T^b\right).
}
This trace is different from zero because the automorphism in the Lie algebra implies $\left[X^a,X^b\right]\orden T^c$. The fact that in this expression appears the commutator between the Goldstone matrix and the broken generators in short implies that $\sigma^b_1T^b$ depends in principle on a linear combination of all the SO(5) generators.  Coming back to eq.~(\ref{sigmaequationexpanded}), separating the term $m=n$ from the rest and taking traces, we obtain a recursive formula for the $n$-th coefficient ($n>1$),
\eq{
2\sigma_n^b T^b=-\sum_{m=1}^{n-1}\sigma^b_m\frac{i^{n-m}}{\left(n-m\right)!}\{T^b,\Pi^{n-m}\}+\alpha^a \frac{i^n}{n!}\left[X^a,\Pi^n\right].
}
In general, the right-handed side will depend on all the SO(5) generators. Using the orthonormal basis of generators, we can multiply both sides of the previous equation by $T^c$ and take the trace to find
\eq{
2\sigma^c_n=\textrm{tr}\Bigg[\Bigg(&-\sum_{m=1}^{n-1}\sigma^b_m\frac{i^{n-m}}{\left(n-m\right)!}\{T^b,\Pi^{n-m}\}
\nn\\&+\alpha^a \frac{i^n}{n!}\Bigg[X^a,\Pi^n\Bigg]\Bigg)T^c\Bigg],\quad n>1.
}
Just for completeness, we also derive $\sigma_2^c$ that turns out to be zero,
\eq{
2\sigma^c_2
&=\textrm{tr}\left[\left(-i\sigma^b_1\{T^b,\Pi\}-\frac{1}{2}\alpha^a\left[X^a,\Pi^2\right]\right)T^c\right]\nn\\
&=\frac{1}{2}\textrm{tr}\left[\left(\alpha^a\textrm{tr}\left(\left[X^a,\Pi\right] T^b\right)\{T^b,\Pi\}-\alpha^a\left[X^a,\Pi^2\right]\right)T^c\right]\nn\\
&=\frac{1}{2}\textrm{tr}\left[\left(\alpha^a\{\left[X^a,\Pi\right],\Pi\}-\alpha^a\left[X^a,\Pi^2\right]\right)T^c\right]\nn\\
&=\frac{1}{2}\textrm{tr}\left[\left(\alpha^a\left[X^a,\Pi^2\right]-\alpha^a\left[X^a,\Pi^2\right]\right)T^c\right]=0,
}
where first we have substituted $\sigma^b_1$ calculated above, 
then we have replaced $A=\tfrac{1}{2}[X^a,\Pi]=\textrm{tr}(AT^b)T^b$ in $\{A,\Pi\}=\textrm{tr}(AT^b)\{T^b,\Pi\}$  and finally we have used $\{\left[X^a,\Pi\right],\Pi\}=\left[X^a,\Pi\right]\Pi+\Pi\left[X^a,\Pi\right]=\left[X^a,\Pi^2\right]$. Therefore, given an infinitesimal SU(5) transformation 
\eq{
V\approx 1+i\alpha^a X^a+i\beta T^b,
}
the corresponding infinitesimal SO(5) transformation reads
\eq{\label{Uequation}
U\approx 1+i\beta^b T^b-\frac{1}{2f}\alpha^a\left[X^a,\Pi\right]+\mathcal{O}\left(\frac{\Pi^3}{f^3}\right),
}
which is a result of the particular embedding of SO(5) into SU(5). 

Now we focus on a gauge transformation, belonging to the subgroup $[\su{2}\times\u{1})]_1\times[\su{2}\times\u{1})]_2\subset\su{5}$ that is spontaneously broken to the SM group $[\su{2}\times\u{1})]$. Then $V_g$ is expanded by no more than the union of the set $\left\{X^a_g\right\}=\left\{Q_1^a-Q_2^a,Y_1-Y_2\right\}$ of broken generators and the set $\left\{T^b_g\right\}=\left\{Q_1^a+Q_2^a,Y_1+Y_2\right\}$ of unbroken ones. However, from eq.~(\ref{Uequation}) it is clear that, due to the non linearity of $U$, restricting ourselves to transformations along the gauge directions in the group does not imply that the matrix $U_g$ depends only on the diagonal gauge subgroup generators: the commutator $\left[X_g^a,\Pi\right]$ cannot be expanded entirely in terms of the SM gauge generators because $\left[X_g^a,\Pi\right]\neq \textrm{tr}\left(\left[X_g^a,\Pi\right]T^b_g\right)T^b_g$, so it requires the full set of SO(5) generators. As a consequence, the Lagrangian $\Lag_{Y_H}^{(b)}$ in eq.~(\ref{lagkappab}) whose second term (the T-parity transformed of the first one) depends explicitly on $\Omega$ (the element of the center of the gauge group, commuting only with the gauge generators) is not invariant under a gauge transformation,
\eq{
 -\kappa f(&\overline{\Psi}_2\xi \Psi_R+\overline{\Psi}_1\Sigma_0\Omega\xi^{\dag}\Omega\Psi_R) \nn\\
&\xrightarrow{G_g}
-\kappa f\left( \overline{\Psi}_2\xi \Psi_R+\overline{\Psi}_1\Sigma_0\Omega\xi^{\dag}U_g^{\dag}\Omega U_g\Psi_R\right),
\label{YchiTodd}
}
because in general $U_g$ does not commute with $\Omega$.

\begin{figure}
\centering
\subfigure[]{\includegraphics[scale=0.6]{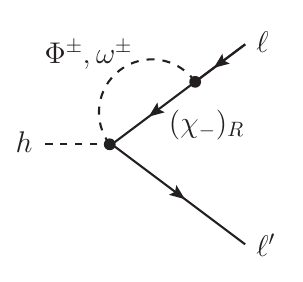}}\qquad
\subfigure[]{\includegraphics[scale=0.6]{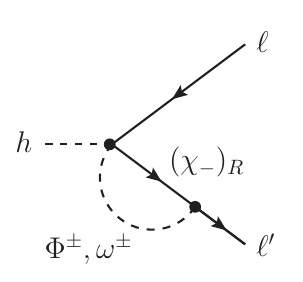}}
\caption{One-loop Feynman diagrams contributing to the ultraviolet divergences of lepton flavor changing Higgs decays in the original LHT if the gauge singlet $\chi$ is T-odd ('t Hooft-Feynman gauge).} \label{fig:diagrams3}
\end{figure}

This has two important implications. First, this implementation of T-parity in the fermionic sector with a {\em $\chi$ T-odd must be discarded} because it is incompatible with gauge invariance. In fact we have found that this Lagrangian gives rise to unmatched divergent contributions in lepton flavor changing Higgs decays \cite{delaguilaInverseSeesawNeutrino2019} from the one-loop diagrams of Fig.~\ref{fig:diagrams3} that involve a $(\chi_-)_R$. On the other hand, regardless of the T-parity realization, it results apparent that the SO(5) multiplets must be complete (as it is the case of $\Psi_R$) because a non-linear gauge transformation $U_g$ mixes all its components and not just those laying in the invariant subspaces under the linear part of (\ref{Uequation}). In particular, the incomplete SO(5) representations $\Psi_L$ and $\Psi_L^\chi$ of eq.~(\ref{PsiL}), introduced to give direct mass terms to $\widetilde{l}^c_-$ and $\chi$ through eq.~(\ref{masswidetildechi}) by coupling them with $\Psi_R$, do not only break the global SO(5) but also the gauge invariance, as we have just shown.

In contrast to the case of a T-odd $\chi_-$ discussed above, the amplitudes for lepton flavor changing Higgs decays are finite at one loop if the gauge singlet is $\chi_+$, T-even. However these amplitudes exhibit a non-decoupling behaviour proportional to the logarithm of the mirror partner masses and to the misalignment between the mass matrices of mirror and mirror-partner leptons \cite{delaguilaLeptonFlavorChanging2017}. This new source of lepton flavor violation can be viewed as a vestige of the broken gauge invariance, that is restored when partners and mirror partners share a complete multiplet and hence get their masses from the same coupling. Actually the one-loop amplitudes are finite thanks to the contributions of both types of T-odd leptons that are individually divergent but cancel each other, so {\em mirror partners cannot be ignored}. Interestingly, the contribution of $\chi_+$ does decouple and is finite by its own.

At any rate, we need a new mechanism to give masses at least of order $f$ to the partner leptons and the $\chi$, because they must be heavy enough to fulfill the EWPD constraints \cite{hubiszPhenomenologyLittlestHiggs2005,hubiszElectroweakPrecisionConstraints2006}. A way to proceed that at the same time is compatible with the gauge symmetry and T-parity is the object of next section.

\section{A gauge-invariant Littlest Higgs with T-parity}\label{new proposal}

In this section we construct in detail the Lagrangian of a new LHT with explicit compatibility between gauge invariance and T-parity in order to address and eventually solve the problems we encountered in previous sections.
The guiding line is the necessity of giving gauge-invariant mass terms to the gauge singlet $\chi_R$ and the mirror-partner leptons $\widetilde{l}^c_R$ without introducing their left-handed counterparts in additional SO(5) multiplets transforming like $\Psi_R$, which would then be incomplete and hence at odds with the gauge symmetry.

\subsection{A minimal extension of the global symmetry}

The simplest way to proceed consists of assuming that the new fermion fields (left-handed components of $\chi$ and $\widetilde{l}^c$) transform only under an external $\su{2}\times \u{1}$. Then they will not mix with the others, as it would happen if they belonged to the same SO(5) multiplet. This requires the enlargement of the original global symmetry group (\ref{extsu5}) to at least:\footnote{In \cite{pappadopuloTparityItsProblems2011} the same extension of the symmetry group was proposed. However they chose to embed the right-handed mirror leptons in a representation of the diagonal subgroup, coupling them to their left-handed counterparts through the new non-linear scalar field. There no mirror partners are introduced, and the fermion SU(2) singlet does not couple to the Higgs hence preventing the unwanted quadratic mass corrections. Here we prefer to keep the fermion content and the structure of Yukawa couplings of the usual LHT, completing the SU(5) fermion multiplets and providing singlet and mirror partners with masses compatible with gauge invariance.
}
\eq{
\textrm{SU(5)}\times \left[\su{2}\times \u{1}\right]''_1\times \left[\su{2}\times \u{1}\right]''_2.
\label{newextsu5}
}
This larger global group gets broken spontaneously when {\it two} non-linear tensor fields, $\Sigma$ and $\widehat\Sigma$, acquire a \vev\ at an energy scale $f$ (for simplicity we take the same scale in both sectors),
\eq{
\textrm{SU(5)}\times[\su{2}&\times \u{1}]''_1\times \left[\su{2}\times \u{1}\right]''_2\nn\\&\xrightarrow{\Sigma_0,\widehat\Sigma_0}\textrm{SO(5)}\times \left[\su{2}\times \u{1}\right]'',
}
where $\Sigma_0$ in eq.~(\ref{Sigma0}) breaks spontaneously SU(5) down to SO(5) as before (see section~\ref{sec:su5toso5}), and $\widehat{\Sigma}_0=\Sigma_0$ breaks the extra piece to its diagonal subgroup $\left[\su{2}\times \u{1}\right]''$, leaving $14+4=18$ Goldstone bosons. This particular breaking direction allows us to take advantage of all the properties already mentioned. On the other hand, since extra \u{1} factors had to be introduced before to accommodate the hypercharges of the right-handed SM charged leptons, this is the minimal and most natural extension one can think of.

Throughout the rest of the work, the notation we follow for the extra fields and their corresponding transformations consists of putting a hat over their symbols. In our construction the new sigma field $\widehat\Sigma$ transforms only under the diagonal subgroup $\su{2}''\times \u{1}''$. To be consistent, for the scalar and fermionic sector we take the representation of $\left[\su{2}\times \u{1}\right]''_1\times \left[\su{2}\times \u{1}\right]''_2$ over a 5-dimensional complex space generated by the same set of matrices of eqs.~(\ref{generators1}) and (\ref{generators2}). 

The gauged subgroup is again $[\su{2}\times \u{1}]_1\times [\su{2}\times \u{1}]_2$ but now it is the sum of the $[\su{2}\times \u{1}]'_1\times [\su{2}\times \u{1}]'_2\subset\textrm{SU(5)}$ plus the extra $[\su{2}\times \u{1}]''_1\times [\su{2}\times \u{1}]''_2$ of eq.~$(\ref{newextsu5})$, so there are the same number of gauge bosons. Likewise, the SM gauge group will be the sum of the $\left[\su{2}\times \u{1}\right]'$ inside SO(5) and the extra $\left[\su{2}\times \u{1}\right]''$. In this way, we can have fermions that transform only under the SM gauge group alleviating the aforementioned difficulties.
The Lagrangian for the gauge fields and their self-interactions is as shown before in eq.~(\ref{lagG}). 

\subsection{The additional Goldstone fields}

According to the SSB, the additional Goldstone matrix is expanded by the broken generators of the extra group $\{Q^a_1-Q^a_2,Y_1-Y_2\}$. It reads
\eq{
\scriptsize
\widehat{\Pi}=\left(\begin{array}{ccccc}-\disp\frac{\widehat{\omega}^0}{2}-\frac{\widehat{\eta}}{\sqrt{20}} & -\disp\frac{\widehat{\omega}^+}{\sqrt{2}} & 0 & 0 & 0 \\
-\disp\frac{\widehat{\omega}^-}{\sqrt{2}} & \disp\frac{\widehat{\omega}^0}{2}-\frac{\widehat{\eta}}{\sqrt{20}} & 0 & 0 & 0 \\
0 & 0 & \sqrt{\disp\frac{4}{5}}\widehat{\eta} & 0 &  0 \\
0 & 0 & 0 & -\disp\frac{\widehat{\omega}^0}{2}-\frac{\widehat{\eta}}{\sqrt{20}} & -\disp\frac{\widehat{\omega}^-}{\sqrt{2}} \\
0 & 0 & 0 & -\disp\frac{\widehat{\omega}^+}{\sqrt{2}} & \disp\frac{\widehat{\omega}^0}{2}-\frac{\widehat{\eta}}{\sqrt{20}}
\end{array}\right).
}
These Goldstone fields are charged only under $[\su{2}\times \u{1}]''_1\times [\su{2}\times \u{1}]''_2$. Under the SM gauge group they decompose in
\eq{
\widehat{\Pi}: 1_0\oplus 3_0,
}
including a new SU(2) triplet
\eq{\widehat{\omega}=\left(\begin{array}{cc}
-\disp\frac{\widehat{\omega}^0}{2} & -\disp\frac{\widehat{\omega}^+}{\sqrt{2}}\\
-\disp\frac{\widehat{\omega}^-}{\sqrt{2}} & \disp\frac{\widehat{\omega}^0}{2}
\end{array}\right)
} 
and a singlet ($\widehat{\eta}$). These scalars have the same quantum numbers as the corresponding unhatted would-be Goldstone bosons (to be eaten by the heavy gauge fields) and will actually mix with them at order $v^2/f^2$. From this matrix we build the non-linear sigma field
\eq{\label{hatxitransformation}
\hat{\xi}=e^{i\widehat{\Pi}/f}\xrightarrow{G''} \widehat{V}\hat{\xi}\widehat{U}^{\dag}=\widehat{U}\hat{\xi}\Sigma_0\widehat{V}^T\Sigma_0,
}
where $\widehat{V}$, $\widehat{U}$ are transformations of the extra group. In the particular case of a gauge transformation $V_g$ and $\widehat{V}_g$ coincide. However, from eqs.~(\ref{Udef}) and (\ref{hatxitransformation}), $U_g$ and $\widehat{U}_g$ are different, since the former depends on $V_g$ and $\Pi$, involving all SO(5) generators, while latter depends on $V_g$ and $\widehat{\Pi}$, requiring just SM generators. 

We also introduce the field that transforms linearly under the extra group,
\eq{
\widehat{\Sigma}=\hat{\xi}\Sigma_0\hat{\xi}^T=\hat{\xi}^2\Sigma_0, \quad \widehat{\Sigma}\xrightarrow{G''} \widehat{V}\widehat{\Sigma}\widehat{V}^T.
}
To assign a T-odd parity to all new scalars we define
\eq{
\widehat{\Pi}\xrightarrow{\textrm{T}} -\widehat{\Pi}
}
and therefore
\eq{
\hat{\xi}\xrightarrow{\textrm{T}} \hat{\xi}^{\dagger},\quad \widehat{\Sigma}\xrightarrow{\textrm{T}} \Sigma_0\widehat{\Sigma}^{\dagger}\Sigma_0.
}

\begin{figure}
\centering
\subfigure[]{\includegraphics[scale=0.6]{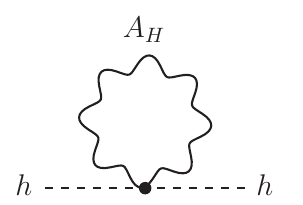}}\qquad
\subfigure[]{\includegraphics[scale=0.6]{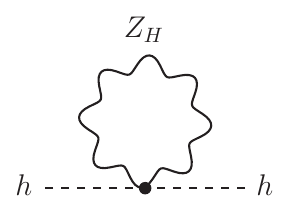}}
\subfigure[]{\includegraphics[scale=0.6]{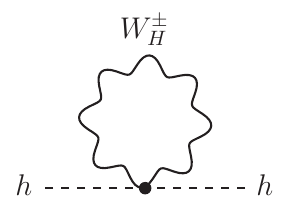}}
\caption{One-loop Feynman diagrams that would lead to unacceptable quadratic divergences in the Higgs mass if the mixed Lagrangian of eq.~(\ref{scalarintercactionlag}) did not vanish.}\label{fig:diagrams4}
\end{figure}

A gauge-invariant and T-parity preserving Lagrangian for the kinetic terms and self-interactions of the new scalars can be built similarly to $\Lag_S$ (\ref{lagS}),
\eq{\label{lagShat}
\Lag_{\widehat{S}}=
\frac{f^2}{8}\textrm{tr}\left[\left(D^{\mu}\widehat{\Sigma}\right)^{\dagger}D_{\mu}\widehat{\Sigma}\right],
}
where the covariant derivative for $\widehat{\Sigma}$ is defined analogously to that in eq.~(\ref{scalarderivative}),
\eq{
D_{\mu}\widehat{\Sigma}&=\partial_{\mu}\widehat{\Sigma}-\sqrt{2}i\sum_{j=1}^2\Big[g W^a_{j\mu}\left(Q^a_j\widehat{\Sigma}+\widehat{\Sigma} Q^{a T}_j\right) \nn\\&-g'B_{j\mu}\left(Y_j\widehat{\Sigma}+\widehat{\Sigma} Y_j^T\right)\Big].
}
At this point one may argue that an additional term mixing both scalar sectors,
\eq{\label{scalarintercactionlag}
\Lag_{S\widehat{S}}=\alpha_{S\widehat{S}}\, f^2 \textrm{tr}\left[\left(D^{\mu}\Sigma\right)^{\dagger}D_{\mu}\widehat{\Sigma}\right]+\hc,
}
is both gauge and T-parity invariant and hence should be included. However this Lagrangian involves couplings of heavy gauge bosons to scalars that lead to unmatched quadratically divergent contributions to the Higgs mass from the diagrams of Fig.~\ref{fig:diagrams4}. Therefore, one must take $\alpha_{S\widehat{S}}=0$.

\subsection{Extra fermions and their interactions}

The original LHT allows for masses of SM and mirror leptons through Yukawa interactions but, in order to be consistent with gauge and T-parity invariance, the right-handed components of mirror leptons must share a complete SO(5) quintuplet $\Psi_R$ with T-odd mirror-partner leptons $(\widetilde{l}^c_-)_R$ and a singlet $(\chi_+)_R$ that must be T-even according to eq.~(\ref{YchiTodd}). To provide these fields with a heavy mass one needs to introduce their left-handed components as well, but they cannot live in SO(5) multiplets transforming like $\Psi_R$ (as the $\Psi_L$ and $\Psi_L^\chi$ of eq.~(\ref{PsiL}) usually introduced) because then they would be incomplete and their vector-like mass terms would break gauge invariance. 

We show below that the goal of giving masses to {\it all} fermions compatible with gauge invariance and T-parity can be achieved in the context of an extended global symmetry with the modified spontaneous breaking pattern described in previous section. In a first proposal we include $(\widetilde{l}^c_-)_L$ and $(\chi_+)_L$ in a quintuplet charged only under the gauged $SU(2)\times U(1)$ subgroup of the enlarged global symmetry, but this minimal model will generate undesired quadratic contributions to the Higgs mass. Then, as a viable solution, we are forced to further extend the fermion content with additional T-even mirror partners $\widetilde{l}^c_+$ and a T-odd $\chi_-$ embedding their left and right handed components in appropriate representations of SU(5) and $\su2\times \u1$, respectively.

\subsubsection{A first attempt introduces quadratic Higgs mass corrections}

Let us introduce the left-handed components of $\widetilde{l}^c$ and $\chi_+$ in such a way that they transform under the SM gauge group but do not mix under an SO(5) transformation. Then, taking a representation of the extra group that acts over the 5-dimensional space, we compose the following multiplet
\eq{
\widehat{\Psi}_L=\left(\begin{array}{c}
-i\sigma^{2}(\widetilde{l}^{c}_{\mathunderscore})_{L}\\
i(\chi_{+})_{L}\\
0
\end{array}\right),\quad \widehat{\Psi}_L\xrightarrow{G_g}\widehat{U}_g\widehat{\Psi}_L
\label{hatPsiL}
}
emphasizing that it transforms non linearly, and not under SO(5) but just under the diagonal subgroup of the extra group. For simplicity, we have chosen that both fields lay in the same multiplet, but they could be split into two and our conclusions would not change. Under the discrete T-parity symmetry we define 
\eq{
\widehat{\Psi}_L\xrightarrow{\textrm{T}}\Omega \widehat{\Psi}_L,
}
in order to assign the proper parities. The right-handed fields form the SO(5) quintuplet $\Psi_R$ of eq.~(\ref{psiR}) with
\eq{
\Psi_{R}=\left(\begin{array}{c}
-i\sigma^{2}(\widetilde{l}_{-}^{c})_R\\
i(\chi_+)_{R}\\
-i\sigma^{2}l_{HR}
\end{array}\right),\quad 
\Psi_R\xrightarrow{G_g} U_g\Psi_R, \quad 
\Psi_R\xrightarrow{\textrm{T}} \Omega \Psi_R.
\label{PsiRTeven}
}

We may now pair $\widehat{\Psi}_L$ with $\Psi_R$ in the following Yukawa Lagrangian
\eq{
\Lag_Y^{\hat{\xi}\xi}=-\kappa' f\overline{\widehat{\Psi}_L}\left(\hat{\xi}^{\dag}\xi+\hat{\xi}\xi^{\dag}\right)\Psi_R+\hc
\label{Laghatxixi}
}
to give $\widetilde{l}^c_-$ and $\chi_+$ a mass of order $\kappa' f$, compatible with gauge invariance and T-parity since
\eq{
\hat{\xi}^{\dag}\xi & \xrightarrow{G_g} 
\widehat{U}_g \hat{\xi}^{\dag}\xi U^{\dag}_g,
\\
\hat{\xi}^\dagger \xi & \xrightarrow{\textrm{T}}
\Omega\hat{\xi}\xi^\dagger\Omega,
}
where we have used that $[\Omega,\hat{\xi}]=0$ because only gauge generators are involved.

\begin{figure}
\centering
\subfigure[]{\includegraphics[scale=0.6]{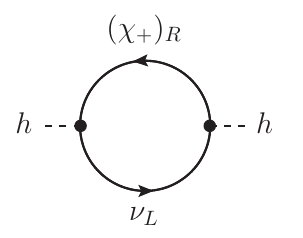}}\qquad
\subfigure[]{\includegraphics[scale=0.6]{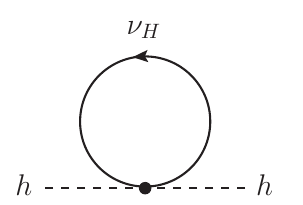}}
\subfigure[]{\includegraphics[scale=0.6]{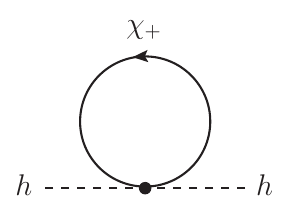}}\qquad
\subfigure[]{\includegraphics[scale=0.6]{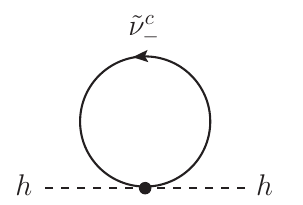}}
\caption{One-loop Feynman diagrams contributing to the quadratic divergences of the Higgs self-energy from the fermion sector in the LHT extended with $\widehat{\Psi}_L$ (\ref{hatPsiL}). Diagrams (a) and (b) from $\Lag_{Y_H}$ cancel each other, whereas (c) and (d) from $\Lag_Y^{\hat{\xi}\xi}$ (\ref{Laghatxixi}) are proportional to $\kappa'^2$ and do {\it not} cancel.} \label{fig:diagrams1}
\end{figure}

Unfortunately, these interactions lead to unacceptable quadratic divergences to the Higgs boson mass from the diagrams in Fig.~\ref{fig:diagrams1}. Diagrams (a) and (b), already in the original LHT, cancel each other.\footnote{The cancellation occurs regardless the mechanism that gives mass to $\chi_+$ because the divergence is independent of the $\chi$ mass \cite{delaguilaInverseSeesawNeutrino2019}.} However (c) and (d), from the new interaction in eq.~(\ref{Laghatxixi}), add up to yield
\eq{
\delta m_h^2=\frac{3\kappa'^2\Lambda^2}{4\pi^2}.
}
In order to prevent such a quadratic divergence, an alternative mechanism is needed to give masses to $\chi$ and the mirror partner leptons.  

\subsubsection{A viable model with consistent fermion representations}\label{definitive proposal}

Instead of including the new left-handed fields in quintuples transforming non linearly under the SM gauge group as above, we proceed to complete the SU(5) multiplets as follows,
\eq{
\Psi_{1}=\left(\begin{array}{c}
-i\sigma^{2}l_{1L}\\
i\chi_{1L}\\
-i\sigma^2\widetilde{l}^c_{1L}
\end{array}\right),\quad \Psi_{2}=\left(\begin{array}{c}
-i\sigma^2\widetilde{l}^c_{2L}\\
i\chi_{2L}\\
-i\sigma^{2}l_{2L}
\end{array}\right),
\label{newsu5multiplets}
}
with the usual
\eq{\label{su5multipletstransformation}
\Psi_1\xrightarrow{G_g} V_g^{*}\Psi_1,\quad \Psi_2\xrightarrow{G_g} V_g\Psi_2,
\quad \Psi_1 \xrightarrow{\textrm{T}}\Omega\Sigma_0 \Psi_2,
}
so that the $(\chi_+)_R$ and the $(\widetilde{l}^c_-)_R$ inside the SO(5) quintuplet $\Psi_R$ of eq.~(\ref{PsiRTeven}) can couple to them and get a mass proportional to $\kappa f$ through the same $\Lag_{Y_H}$ in eq.~(\ref{lagkappaa}), like the mirror leptons $l_H$ do. 

But still the T-odd combination $(\chi_{\mathunderscore})_{L}=(\chi_{1L}-\chi_{2L})/\sqrt{2}$ and the T-even combination of the mirror partners $(\widetilde{l}^c_{+})_L=(\widetilde{l}^c_{1L}-\widetilde{l}^c_{2L})/\sqrt{2}$ remain massless. (Note that although the T-parities of these fields are different the relative sign in previous definitions is the same according to eq.~(\ref{Tevenchi}).) At this stage, barring the explicit breaking due to the gauge interactions of $\left[\su{2}\times \u{1}\right]^2$, the theory would be SU(5) invariant because now the fermion multiplets are complete, so the Higgs would be an exact Goldstone boson with no mass corrections from this sector through loops at any order in perturbation theory. To give a mass to these additional combinations of fields, we introduce their corresponding right-handed components in an incomplete multiplet that only transforms under the SM gauge group (analogous to eq.~(\ref{hatPsiL}) but for opposite chiralities and T-parities),
\eq{\label{newmultiplet}
\widehat{\Psi}_{R}=\left(\begin{array}{c}
-i\sigma^{2}(\widetilde{l}_{+}^{c})_R\\
i(\chi_{\mathunderscore})_{R}\\
0
\end{array}\right),\quad \widehat{\Psi}_{R}\xrightarrow{G_g} \widehat{U}_g\widehat{\Psi}_R, \quad \widehat{\Psi}_{R}\xrightarrow{\textrm{T}} -\Omega\widehat{\Psi}_R,
}
where we emphasize one more time that the new T-even partner lepton doublet and the new T-odd singlet do not mix under a gauge group transformation what allows them to be separated in different SO(5) multiplets. Finally, we couple this multiplet to $\Psi_1$ and $\Psi_2$ through the non linear field $\hat\xi$,
\eq{
\Lag_{\widehat{Y}_H}=-\widehat{\kappa} f\left(\overline{\Psi}_2\hat{\xi}-\overline{\Psi}_1\Sigma_0\hat{\xi}^{\dag}\right)\widehat{\Psi}_R+\hc,\label{Lagkappahat}
}
using again that $\Omega$ commutes with $\hat{\xi}$. This way $\widetilde{l}^c_+$ and $\chi_-$ get a mass of order $\widehat\kappa f$.

\begin{figure}
\centering
\subfigure[]{\includegraphics[scale=0.6]{./hh-chiTevennu.pdf}}\qquad
\subfigure[]{\includegraphics[scale=0.6]{./hh-nuH.pdf}}
\subfigure[]{\includegraphics[scale=0.6]{./hh-chiTeven.pdf}}\qquad
\subfigure[]{\includegraphics[scale=0.6]{./hh-nucTodd.pdf}}
\subfigure[]{\includegraphics[scale=0.6]{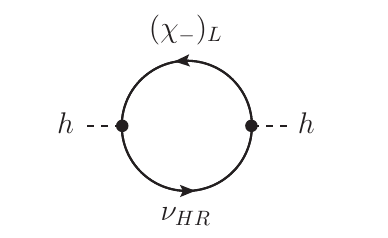}}\qquad
\subfigure[]{\includegraphics[scale=0.6]{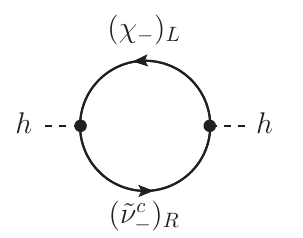}}
\subfigure[]{\includegraphics[scale=0.6]{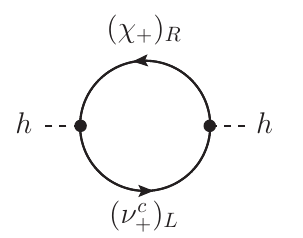}}\label{diag27}
\caption{One-loop Feynman diagrams contributing to the quadratic divergences of the Higgs self-energy from the fermion sector in the new LHT. Diagrams (a) and (b) arise from $\Lag_{Y_H}$ and cancel each other, as in Fig.~\ref{fig:diagrams1}. The rest stem from $\Lag_{\widehat{Y}_H}$ (\ref{Lagkappahat}) and also cancel among themselves.} \label{fig:diagrams2}
\end{figure}

As this new sector does not have a direct coupling with the Higgs field and the Lagrangian in eq.~(\ref{lagkappaa}) with complete multiplets is SU(5) invariant, we do not expect quadratic divergences in the Higgs mass. 
In fact, the relevant one-loop diagrams, shown in Fig.~\ref{fig:diagrams2}, result in a logaritmically divergent correction to the Higgs mass of
\eq{\label{higgs mass}
\delta m^2_h=\frac{3f^2\kappa^2\widehat{\kappa}^2}{4\pi^2}\log\Lambda^2.
} 
This result obtained diagrammatically will be reproduced in section~\ref{BFM} from the calculation of the Coleman-Weinberg potential using the background field method.

There remains the introduction of kinetic terms and gauge interactions of all fermion fields in the model.
The new fields $\widetilde{l}^c_{rL}$ and $\chi_{rL}$ that make up the left-handed components of $\widetilde{l}^c_\pm$ and $\chi_\pm$ belong to the SU(5) quintuplets $\Psi_r$ ($r=1,2$) so they will get their kinetic terms and gauge interactions from the same Lagrangian $\Lag_{F_L}$ of eq.~(\ref{LagFL}). The right-handed components of $\widetilde{l}^c_-$ and $\chi_+$ were already in the SO(5) quintuplet $\Psi_R$ with kinetic terms and interactions from $\Lag_{F_R}$ in eq.~(\ref{LagFR}). For the right-handed fields $(\widetilde{l}^c_+)_R$ and $(\chi_-)_R$ in the new SO(5) quintuplet $\widehat{\Psi}_R$ we introduce 
\eq{
\Lag_{\widehat{F}_R}=
i\overline{\widehat{\Psi}}_R\gamma^{\mu}\left[\partial_{\mu}+\frac{1}{2}\hat{\xi}^{\dag}\left(D_{\mu}\hat{\xi}\right)+\frac{1}{2}\hat{\xi}\Sigma_0D^{*}_{\mu}\left(\Sigma_0\hat{\xi}^{\dag}\right)\right]\widehat{\Psi}_R,
}
with the covariant derivative in eq.~(\ref{covderfer}). Notice that under a T-parity transformation
\eq{
\overline{\widehat{\Psi}}_R\hat{\xi}^{\dag}D_{\mu}\hat{\xi}\widehat{\Psi}_R\xrightarrow{\textrm{T}} \overline{\widehat{\Psi}}_R\hat{\xi}\Sigma_0D^*_{\mu}\left(\Sigma_0\hat{\xi}^{\dag}\right)\widehat{\Psi}_R,
}
since $\Omega$ commutes with $\hat{\xi}$ and the gauge generators. The SM right-handed leptons are singlets under $\su{2}_1''\times\su{2}_2''$ and only the charged ones get their proper hypercharge under the extra $\u{1}_1''\times\u{1}_2''$ with gauge interactions $\Lag_{F'}$ already given in eq.~(\ref{LagFprime}).

For the sake of clarity, we show in tables~\ref{table:hypercharge3} and \ref{table:hypercharge4} the transformation properties of the fermion fields under the different $\su{2}\times\u{1}$ factors. The fields belonging to the complete SO(5) representation are written in their corresponding multiplet in order to emphasize that they mix under a gauge transformation.

\begin{table*}
\caption{Charge assignments under the different $\left[\su{2}\times \u{1}\right]_1\times\left[\su{2}\times \u{1}\right]_2$.}\label{table:hypercharge3}
\centering
\begin{tabular}{|c|c|c|c|}
\hline
 & \quad $\left[\su{2}'\times \u{1}'\right]^2\subset \textrm{SU(5)}$ \quad & \quad $\left[\su{2}''\times \u{1}''\right]^2$ \quad & \quad $\left[\su{2}\times \u{1}\right]^2_{\textrm{gauge}}$   \quad \T\B\\ 
\hline
$l_{2L}$ & $\left(1,2\right)_{\left(-\frac{1}{5},-\frac{3}{10}\right)}$ & $\left(1,1\right)_{\left(0,0\right)}$ & $\left(1,2\right)_{\left(-\frac{1}{5},-\frac{3}{10}\right)}$   \T\B\\ 
$l_{1L}$ & $\left(2,1\right)_{\left(-\frac{3}{10},-\frac{1}{5}\right)}$ & $\left(1,1\right)_{\left(0,0\right)}$ & $\left(2,1\right)_{\left(-\frac{3}{10},-\frac{1}{5}\right)}$ \T\B\\ 
$\chi_{2L}$ & $\left(1,1\right)_{\left(-\frac{1}{5},\frac{1}{5}\right)}$ & $\left(1,1\right)_{\left(0,0\right)}$ & $\left(1,1\right)_{\left(-\frac{1}{5},\frac{1}{5}\right)}$ \T\B\\
$\chi_{1L}$ & $\left(1,1\right)_{\left(\frac{1}{5},-\frac{1}{5}\right)}$ & $\left(1,1\right)_{\left(0,0\right)}$ & $\left(1,1\right)_{\left(\frac{1}{5},-\frac{1}{5}\right)}$   \T\B\\
$\widetilde{l}^c_{2L}$ &  $\left(2,1\right)_{\left(\frac{3}{10},\frac{1}{5}\right)}$ &  $\left(1,1\right)_{\left(0,0\right)}$ &  $\left(2,1\right)_{\left(\frac{3}{10},\frac{1}{5}\right)}$    \T\B\\
$\widetilde{l}^c_{1L}$ &  $\left(1,2\right)_{\left(\frac{1}{5},\frac{3}{10}\right)}$ & $\left(1,1\right)_{\left(0,0\right)}$ & $\left(1,2\right)_{\left(\frac{1}{5},\frac{3}{10}\right)}$  \T\B\\
$\ell_{R}$ &  $\left(1,1\right)_{\left(0,0\right)}$ & $\left(1,1\right)_{\left(-\frac{1}{2},-\frac{1}{2}\right)}$ & $\left(1,1\right)_{\left(-\frac{1}{2},-\frac{1}{2}\right)}$  \T\B\\
$X$ &  $\left(1,1\right)_{\left(\frac{3}{5},\frac{2}{5}\right)}$ & $\left(1,1\right)_{\left(-\frac{1}{2},-\frac{1}{2}\right)}$ & $\left(1,1\right)_{\left(\frac{1}{10},-\frac{1}{10}\right)}$  \T\B\\
$X^{*}$ &  $\left(1,1\right)_{\left(\frac{2}{5},\frac{3}{5}\right)}$ & $\left(1,1\right)_{\left(-\frac{1}{2},-\frac{1}{2}\right)}$ & $\left(1,1\right)_{\left(-\frac{1}{10},\frac{1}{10}\right)}$  \T\B\\[1ex]
\hline
\end{tabular}
\end{table*}

\begin{table*}
\caption{Charge assignments under the different $\left[\su{2}\times \u{1}\right]$.}\label{table:hypercharge4}
\centering
\begin{tabular}{|c|c|c|c|}
\hline
 &  $\su{2}'\times \u{1}'\subset \textrm{SO(5)} $  &  $\su{2}''\times \u{1}''$  &  $\left[\su{2}\times \u{1}\right]_{\textrm{gauge}}$    \T\B\\ 
\hline
$\Psi_{R}=\left(\begin{array}{c}
-i\sigma^{2}(\widetilde{l}^{c}_{\mathunderscore})_{R}\\
i(\chi_{+})_{R}\\
-i\sigma^{2}l_{HR}
\end{array}\right)$ & $\left(\begin{array}{c}
2_{\frac{1}{2}}\\
1_0\\
2_{-\frac{1}{2}}
\end{array}\right)$ & $1_{0}$ & $\left(\begin{array}{c}
2_{\frac{1}{2}}\\
1_0\\
2_{-\frac{1}{2}}
\end{array}\right)$  \T\B\\
$(\widetilde{l}^c_{+})_R$ &  $1_0$ &  $2_{\frac{1}{2}}$ &  $2_{\frac{1}{2}}$    \T\B\\
$(\chi_{\mathunderscore})_{R}$ & $1_0$ & $1_0$ & $1_0$   \T\B\\
\hline
\end{tabular}
\end{table*}

To summarize, the full Lagrangian of the new LHT model reads:
\eq{
\Lag &=
  \Lag_G + \Lag_S + \Lag_{\widehat{S}} 
+ \Lag_{F_L} + \Lag_{F_R} + \Lag_{\widehat{F}_R} + \Lag_{F'}
\nn\\&+ \Lag_{Y_H} + \Lag_{\widehat{Y}_H} + \Lag_Y.
}

\subsection{Physical fields}

\subsubsection{Gauge bosons}

After the electroweak SSB, the SM gauge bosons are obtained from the T-even fields of eq.~(\ref{gaugeTeven}) by diagonalizing the Lagrangian $\Lag_S$ of eq~(\ref{lagS}),
\eq{
W^{\pm}=\frac{1}{\sqrt{2}}\left(W^1\mp i W^2\right),\quad \left(\begin{array}{c}
Z\\
A
\end{array}\right)=\left(\begin{array}{cc}
c_{W} & s_{W}\\
-s_{W} & c_{W}
\end{array}\right)\left(\begin{array}{c}
W^{3}\\
B
\end{array}\right)
}
with
\eq{
W^a=\frac{W^a_1+W^a_2}{\sqrt{2}},\quad B=\frac{B_1+B_2}{\sqrt{2}}.
}
To get the T-odd gauge bosons, one needs to expand both $\Lag_S$ and $\Lag_{\widehat{S}}$ (\ref{lagShat}) up to order $v^2/f^2$ to derive the heavy physical fields from those in eq.~(\ref{gaugeTodd}),
\eq{
W^{\pm}_H&=\frac{1}{\sqrt{2}}\left(W^1_H\mp i W^2_H\right),\quad 
\\
\left(\begin{array}{c}
Z_{H}\\
A_{H}
\end{array}\right)&=\left(\begin{array}{cc}
1 & -x_{H}\frac{v^{2}}{f^{2}}\\
x_{H}\frac{v^{2}}{f^{2}} & 1
\end{array}\right)\left(\begin{array}{c}
W_{H}^{3}\\
B_{H}
\end{array}\right)
}
with
\eq{
W^a_H=\frac{W^a_1-W^a_2}{\sqrt{2}},\quad B_H=\frac{B_1-B_2}{\sqrt{2}},\quad x_H=\frac{5 g g'}{8\left(5g^2-g'^2\right)}.
}
Their corresponding masses to order $v^2/f^2$ are
\eq{
M_W&=\frac{g v}{2}\left(1-\frac{v^2}{12f^2}\right),\quad M_Z=M_W/c_W,\quad \\
M_{W_H}&=M_{Z_H}=\sqrt{2}g f\left(1-\frac{v^2}{16f^2}\right),\quad \\ M_{A_H}&=\sqrt{\frac{2}{5}}g' f\left(1-\frac{5v^2}{16f^2}\right).
}
Notice that even though the gauge bosons are the same as in the original model, the masses of the T-odd combinations are at leading order a factor of $\sqrt{2}$ heavier (see for instance \cite{delaguilaLeptonFlavorChanging2017}). This is because the new extra scalar sector parametrized by $\widehat{\Sigma}$ also takes the \vev\ $\Sigma_0$ hence giving an additional contribution to the heavy gauge boson masses. However the T-even gauge bosons couple only to the Higgs field, belonging to $\Sigma$, and higher order corrections are forbidden by T-parity, so their masses do not change.

\subsubsection{Scalars after gauge fixing}

The spontaneous breaking of gauge symmetries leads to kinetic mixing between gauge bosons and would-be Goldstone boson fields. In the mass eigenbasis, these unwanted mixing terms can be removed, up to an irrelevant total derivative, by introducing the appropriate gauge-fixing Lagrangian
\eq{
\Lag_{\textrm{gf}} &= 
 -\frac{1}{2\xi_\gamma}(\d_\mu A^\mu)^2
 -\frac{1}{2\xi_Z}(\d_\mu Z^\mu-\xi_Z M_Z \pi^0)^2 \nn\\
&-\frac{1}{\xi_W}|\d_\mu W^\mu+i\xi_W M_W \pi^-|^2
 -\frac{1}{2\xi_{A_H}}(\d_\mu A_H^\mu+\xi_{A_H} M_{A_H} \eta)^2 \nn\\
&-\frac{1}{2\xi_{Z_H}}(\d_\mu Z_H^\mu-\xi_{Z_H} M_{Z_H} \omega^0)^2 \nn\\
&-\frac{1}{\xi_{W_H}}|\d_\mu W_H^\mu+i\xi_{W_H} M_{W_H} \omega^-|^2,
\label{eq:gf}
}
defining which Goldstone fields are unphysical and can be absorbed.

After the SSB, the kinetic terms of the scalar fields we have introduced are neither diagonal nor canonically normalized. Besides, the new set of T-odd scalars from the extra group mix with the old ones that have the same quantum numbers. In order to define the physical scalars and identify the actual would-be-Goldstone fields we will perform the following redefinitions in two steps. First we perform a rotation of $45^{\circ}$ in the subspace of every scalar pair with same quantum numbers, so that only one of them (the unhatted) will retain the kinetic mixing with a gauge boson hence becoming the actual would-be-Goldstone field at leading order,
\eq{
\omega^{\pm}\rightarrow
\frac{1}{\sqrt{2}}\left(\omega^{\pm}-\widehat{\omega}^{\pm}\right),\quad & \widehat{\omega}^{\pm}\rightarrow
\frac{1}{\sqrt{2}}\left(\omega^{\pm}+\widehat{\omega}^{\pm}\right), \\
\omega^{0}\rightarrow
\frac{1}{\sqrt{2}}\left(\omega^{0}-\widehat{\omega}^{0}\right),\quad & \widehat{\omega}^{0}\rightarrow
\frac{1}{\sqrt{2}}\left(\omega^{0}+\widehat{\omega}^{0}\right), \\
\eta\rightarrow\frac{1}{\sqrt{2}}\left(\eta-\widehat{\eta}\right),\quad & \widehat{\eta}\rightarrow
\frac{1}{\sqrt{2}}\left(\eta+\widehat{\eta}\right).
}
At this point all kinetic-mixing terms are of order $v^2/f^2$. In the next step we impose that all kinetic terms are canonically normalized and diagonal so that the actual would-be-Goldstone fields could still be removed by the gauge fixing (\ref{eq:gf}) at order $v^2/f^2$. To that end we rescale and mix them as follows,\footnote{We follow the same procedure as in \cite{hubiszElectroweakPrecisionConstraints2006} for the original LHT, but the shifts are different in our model.}
\eq{
h & \rightarrow h,\\
\pi^0 & \rightarrow \pi^0\left(1+\frac{v^2}{12 f^2}\right),\\
\pi^{\pm} & \rightarrow \pi^{\pm}\left(1+\frac{v^2}{12 f^2}\right),\\
\Phi^0 & \rightarrow \Phi^0\left(1+\frac{v^2}{12f^2}\right),\\
\Phi^P & \rightarrow \Phi^P\left(1+\frac{v^2}{12f^2}\right) \nn\\
&\quad+\left(-\omega^0-\widehat{\omega}^0+\sqrt{5}\left(\eta+\widehat{\eta}\right)\right)\frac{v^2}{12f^2},\\
\Phi^{\pm} & \rightarrow \Phi^{\pm}\left(1+\frac{v^2}{24f^2}\right)\pm i\left(\omega^{\pm}+\widehat{\omega}^{\pm}\right)\frac{v^2}{12\sqrt{2}f^2},\\
\Phi^{\pm\pm}&\rightarrow \Phi^{\pm\pm},\\
\eta & \rightarrow \eta \left(1+\frac{5v^2}{48f^2}\right) \nn\\
&\quad+\frac{-5 g' \widehat{\eta}-\sqrt{5}\left[g'\left(\omega^0-\widehat{\omega}^0+2\Phi^P\right)-12 g x_H \omega^0\right]}{12 g'}\frac{v^2}{f^2},\\
\omega^0 & \rightarrow \omega^0\left(1+\frac{v^2}{48f^2}\right)\nn\\
&\quad+\frac{5g\left(-\widehat{\omega}^0+2\Phi^P+\sqrt{5}\widehat{\eta}\right)-\sqrt{5}\left(5 g +12 g' x_H\right)\eta}{60g}\frac{v^2}{f^2},\\
\omega^{\pm} & \rightarrow \omega^{\pm}\left(1+\frac{v^2}{48f^2}\right)+\left(\pm i\sqrt{2}\Phi^{\pm}-\widehat{\omega}^{\pm}\right)\frac{v^2}{12f^2}, \\
\widehat{\eta} & \rightarrow \widehat{\eta}\left(1+\frac{5v^2}{48f^2}\right)+\left(5\eta-\sqrt{5}\omega^0\right)\frac{v^2}{24f^2},\\
\widehat{\omega}^0 & \rightarrow  \widehat{\omega}^0\left(1+\frac{v^2}{48f^2}\right)+\left(\omega^0-\sqrt{5}\widehat{\eta}-\sqrt{5}\eta\right)\frac{v^2}{24f^2},\\
\widehat{\omega}^{\pm} & \rightarrow \widehat{\omega}^{\pm}\left(1+\frac{v^2}{48f^2}\right)+\omega^{\pm}\frac{v^2}{24f^2}.
}
After these redefinitions, the scalars $\eta$, $\omega^0$ and $\omega^{\pm}$ are the would-be-Goldstone bosons of the SSB of the gauge group down to the SM gauge group, eaten by $A_H$, $Z_H$ and $W^{\pm}_H$. Similarly, $\pi^0$ and $\pi^{\pm}$ are the would-be-Goldstone bosons of the SSB of the SM gauge group down to $\u{1}_\textrm{em}$, eaten by $Z$ and $W^{\pm}$. The remaining scalar fields are all physical. They are the Higgs boson, a triplet of hypercharge $Y=1$ composed of $\Phi^{\pm\pm}$, $\Phi^{\pm}$, $\Phi^0$ and $\Phi^P$, and the four new scalars of hypercharge $Y=0$, a singlet $\widehat{\eta}$ and a triplet composed of $\widehat{\omega}^0$ and $\widehat{\omega}^{\pm}$. All of them get a mass by gauge and Yukawa interactions from the Coleman-Weinberg potential \cite{colemanRadiativeCorrectionsOrigin1973,hanNeutrinoMassesLeptonnumber2005}: the triplet $\Phi$ receives a mass of order $f$ from quadratic contributions and the rest from logarithmic contributions to the potential. As a consequence the masses of $\widehat{\eta}$, $\widehat{\omega}^0$ and $\widehat{\omega}^{\pm}$ are independent of $f$, but they can still be large thanks to the interplay of different Yukawa couplings (see subsection~\ref{BFM-physcalars}).

\subsubsection{Fermion masses and mixings}

When the fermion content of the model is extended beyond one family, the Yukawa couplings $\kappa$, $\widehat{\kappa}$ and $\lambda$ in $\Lag_{Y_H}$, $\Lag_{\widehat{Y}_H}$ and $\Lag_Y$, respectively, must be understood as $3\times3$ matrices in flavor space. Omitting flavor indices, for each of the three SM (T-even) left-handed lepton doublets ($l_L$) there is a vector-like doublet of heavy T-odd mirror leptons ($l_H$),
\eq{
l_L&=\left(\begin{array}{c}
\nu_L\\
\ell_L
\end{array}\right)=\frac{l_{1L}-l_{2L}}{\sqrt{2}},\quad \\\
l_{HL}&=
\left(\begin{array}{c}
\nu_{HL}\\
\ell_{HL}
\end{array}\right)=\frac{l_{1L}+l_{2L}}{\sqrt{2}},\quad 
l_{HR}=\left(\begin{array}{c}
\nu_{HR}\\
\ell_{HR}
\end{array}\right),
}
where
\eq{
l_{rL}=\left(\begin{array}{c}
\nu_{rL}\\
\ell_{rL}
\end{array}\right),\quad r=1,2
}
are part of the SU(5) multiplets $\Psi_r$ in eq.~(\ref{newsu5multiplets}) and $l_{HR}$ is part of the SO(5) multiplet $\Psi_R$ in eq.~(\ref{psiR}). The SM right-handed charged leptons $\ell_R$ are singlets under the full SU(5) but have hypercharges under the external $\u{1}''$ groups.

In addition there are two heavy right-handed mirror-partner doublets,
\eq{
(\widetilde{l}^c_{\mathunderscore})_{R}=\left(\begin{array}{c}
(\widetilde{\nu}^c_{\mathunderscore})_{R}\\
(\widetilde{\ell}^c_{\mathunderscore})_{R}
\end{array}\right),\quad (\widetilde{l}^c_{+})_{R}=\left(\begin{array}{c}
(\widetilde{\nu}^c_{+})_{R}\\
(\widetilde{\ell}^c_{+})_{R}
\end{array}\right).
}
The first one is T-odd and is necessary to complete the SO(5) multiplet $\Psi_R$ together with $(\chi_+)_R$, while the second is T-even and lives in the incomplete multiplet $\widehat{\Psi}_R$ in eq.~(\ref{newmultiplet}) along with $(\chi_{\mathunderscore})_R$, charged only under the external $\su{2}''\times \u{1}''$. Their corresponding left-handed counterparts come from the SU(5) multiplets, 
\eq{
(\widetilde{l}^c_{\mathunderscore})_L
&=\left(\begin{array}{c}
(\widetilde{\nu}^c_{\mathunderscore})_{L},\\
(\widetilde{\ell}^c_{\mathunderscore})_{L}
\end{array}\right)
=\frac{(\widetilde{l}^c_1)_L+(\widetilde{l}^c_2)_L}{\sqrt{2}}
,\quad \\
(\widetilde{l}^c_{+})_L
&=\left(\begin{array}{c}
(\widetilde{\nu}^c_{+})_{L},\\
(\widetilde{\ell}^c_{+})_{L}
\end{array}\right)
=\frac{(\widetilde{l}^c_1)_L-(\widetilde{l}^c_2)_L}{\sqrt{2}}.
}
with
\eq{
(\widetilde{l}^c_r)_L=\left(\begin{array}{c}
(\widetilde{\nu}^c_r)_L\\
(\widetilde{\ell}^c_r)_L
\end{array}\right),\quad r=1,2.
}

Finally we have the aforementioned $(\chi_{+})_R$ and $(\chi_{-})_R$. Their left-handed counterparts are the combinations with proper T-parities of the fields $\chi_{1L}$ and $\chi_{2L}$ completing the SU(5) multiplets,
\eq{
(\chi_{+})_L=\frac{\chi_{1L}+\chi_{2L}}{\sqrt{2}},\quad (\chi_{\mathunderscore})_L=\frac{\chi_{1L}-\chi_{2L}}{\sqrt{2}}.
}

Next we introduce flavor indices and derive the mass eigenstates. Since T-parity is exact, the SM (T-even) charged leptons $\ell$ do not mix with the heavy T-odd combinations. They cannot mix with the T-even $\widetilde{\ell}^c_{+}$ either, because they have opposite hypercharge and a coupling through the Yukawa Lagrangian $\Lag_{\widehat{Y}_H}$ would require a non-existing scalar field of hypercharge $Y=1$ acquiring a \vev. Therefore, the SM mass eigenstates result from the diagonalization of the matrix $\lambda$ in eq.~(\ref{laglambda}) leading to the replacements
\eq{
\ell_{L}\rightarrow V^{\ell}_L \ell_{L} ,\quad
\ell_{R}\rightarrow V^{\ell}_R \ell_{R},
}
and the tree-level SM charged lepton masses from
\eq{
\frac{\lambda v}{\sqrt{2}}\left(1-\frac{v^2}{12f^2}\right)= V^{\ell}_L m_{\ell}V^{\ell\dag}_R,
}
where $V^{\ell}_{L,R}$ are unitary matrices in flavor space. Likewise, the heavy charged lepton mass eigenstates are obtained by the replacements
\eq{
\ell_{HL}\rightarrow V^{H}_{L} \ell_{HL} &,\quad
\ell_{HR}\rightarrow V^{H}_{R} \ell_{HR},\\
(\widetilde{\ell}^c_{\mathunderscore})_{L}\rightarrow V^{H}_{L} (\widetilde{\ell}^c_{\mathunderscore})_{L} &,\quad
(\widetilde{\ell}^c_{\mathunderscore})_{R}\rightarrow V^{H}_{R} (\widetilde{\ell}^c_{\mathunderscore})_{R},\\
(\widetilde{\ell}^c_{+})_{L}\rightarrow V^{\widetilde{\ell}^c_{+}}_{L} (\widetilde{\ell}^c_{+})_{L} &,\quad
(\widetilde{\ell}^c_{+})_{R}\rightarrow V^{\widetilde{\ell}^c_{+}}_{R} (\widetilde{\ell}^c_{+})_{R},
}
with 
\eq{
&\sqrt{2}\kappa f= V^{H}_{L} m_{\ell_H}V^{H\dag}_{R}= V^{H}_{L} m_{\widetilde{\ell}^c_{\mathunderscore}}V^{H\dag}_{R},\\
&\sqrt{2}\widehat{\kappa} f= V^{\widetilde{\ell}^c_{+}}_{L} m_{\widetilde{\ell}^c_{+}}V^{\widetilde{\ell}^c_{+}\dag}_{R},
}
where $V^{H}_{L,R}$ and $V^{\widetilde{\ell}^c_{+}}_{L,R}$ are unitary matrices. In contrast to \cite{delaguilaLeptonFlavorChanging2017}, in our model the T-odd mirror lepton doublets $l_H$ and their partners $\widetilde{l}^c_{-}$ rotate with the same matrix as the T-even singlets $\chi_+$, getting a mass proportional to $\kappa$ from the Yukawa Lagrangian $\Lag_{Y_H}$. The new combinations with opposite T-parities, $\widetilde{l}^c_{+}$ and $\chi_{\mathunderscore}$, get masses proportional to $\widehat{\kappa}$ from $\Lag_{\widehat{Y}_H}$. 
Then for the neutral lepton sector the fields have to be redefined as follows,\footnote{%
In \cite{delaguilaLeptonFlavorChanging2017} the partner lepton fields $\widetilde{l}$ are rotated with matrices $V_{L,R}^{\widetilde\ell}$. Here we adopt the convention of rotating their conjugates $\widetilde{l}^c$, which seems more natural as these are the ones embedded in the SO(5) quintuplet. To relate both conventions, $V^{\widetilde{\ell}}_L\equiv \left(V^{\widetilde{\ell}^c_{\mathunderscore}}_R\right)^{*}$ and $V^{\widetilde{\ell}}_R\equiv \left(V^{\widetilde{\ell}^c_{\mathunderscore}}_L\right)^{*}$.}
\eq{
\nu_{HL}\rightarrow V^{H}_{L} \nu_{HL} &,\quad
\nu_{HR}\rightarrow V^{H}_{R} \nu_{HR},\\
(\widetilde{\nu}^c_{\mathunderscore})_{L}\rightarrow V^{H}_{L} (\widetilde{\nu}^c_{\mathunderscore})_{L} &,\quad
(\widetilde{\nu}^c_{\mathunderscore})_{R}\rightarrow V^{H}_{R} (\widetilde{\nu}^c_{\mathunderscore})_{R},\\
(\widetilde{\nu}^c_{+})_{L}\rightarrow V^{\widetilde{\ell}^c_{+}}_{L} (\widetilde{\nu}^c_{+})_{L} &,\quad
(\widetilde{\nu}^c_{+})_{R}\rightarrow V^{\widetilde{\ell}^c_{+}}_{R} (\widetilde{\nu}^c_{+})_{R},\\
(\chi_{+})_{L}\rightarrow V^{H}_{L} (\chi_{+})_{L} &,\quad
(\chi_{+})_{R}\rightarrow V^{H}_{R}(\chi_{+})_{R},\\
(\chi_{\mathunderscore})_{L}\rightarrow  V^{\widetilde{\ell}^c_{+}}_{L} (\chi_{\mathunderscore})_{L} &,\quad
(\chi_{\mathunderscore})_{R}\rightarrow  V^{\widetilde{\ell}^c_{+}}_{R} (\chi_{\mathunderscore})_{R}
}
and the corresponding mass matrices are diagonalized by
\eq{\label{neutralmassmatrices}
\sqrt{2}\kappa f \left(1-\frac{v^2}{8f^2}\right)&=V^{H}_{L} m_{\nu_H}V^{H\dag}_{R}=V^{H}_{L} m_{\widetilde{\nu}^c_{\mathunderscore}}V^{H\dag}_{R},\\
\sqrt{2}\kappa f \left(1-\frac{v^2}{4f^2}\right)&=V^{H}_{L} m_{\chi_{+}}V^{H\dag}_{R},\\
\sqrt{2}\widehat{\kappa} f &=V^{\widetilde{\ell}^c_{+}}_{L} m_{\chi_{\mathunderscore}}V^{\widetilde{\ell}^c_{+}\dag}_{R}= V^{\widetilde{\ell}^c_{+}}_{L} m_{\widetilde{\nu}^c_{+}}V^{\widetilde{\ell}^c_{+}\dag}_{R}.
}
To find the mass eigenstates of the neutral leptons one also has to take into account that those with same gauge quantum numbers mix as well when the Lagrangian is expanded up to order $v^2/f^2$ (see table~\ref{table:mixing}). The mixing of order $v/f$ is the most pressing to include, since it corrects the masses at order $v^2/f^2$. The mixing of order $v^2/f^2$ only plays a role in the diagonalization matrix because it enters in the masses at order $v^4/f^4$.

\begin{table*}
\caption{Order of the mixing between neutral fields. A dot means that they are connected by the mass term and a dash indicates that no mixing term is generated to order $v^2$. }\label{table:mixing}
\centering
\begin{tabular}{|c|c|c|c|c|c|c|}
\hline
 & \quad $(\chi_{+})_L$ \quad & \quad $(\chi_{\mathunderscore})_L$ \quad & \quad $\nu_L$   \quad  & \quad $\nu_{HL}$   \quad  & \quad $(\widetilde{\nu}^c_{+})_L$   \quad  & \quad $(\widetilde{\nu}^c_{\mathunderscore})_L$   \quad    \T\B\\ 
\hline
$(\chi_{+})_R$ & $\bullet$ & \textendash & $\orden v $ & \textendash & $\orden v $ & \textendash     \T\B\\ 
$(\chi_{\mathunderscore})_R$ & \textendash   & $\bullet$ & \textendash & \textendash & \textendash  & \textendash  \T\B\\ 
$\nu_{HR}$ & \textendash  & $\orden v$ & \textendash & $\bullet$ & \textendash & $\orden v^2$  \T\B\\
$(\widetilde{\nu}^c_{+})_R$ & \textendash &  \textendash &  \textendash & \textendash & $\bullet$ & \textendash \T\B\\
$(\widetilde{\nu}^c_{\mathunderscore})_R$ &  \textendash & $\orden v$ & \textendash & $\orden v^2$ & \textendash & $\bullet$  \T\B\\
\hline
\end{tabular}
\end{table*}

The misalignment between the sectors $\kappa$, $\widehat{\kappa}$ and $\lambda$ is a source of flavor mixing. The flavor mixing matrices parametrizing this misalignment can be defined as follows:
\eq{
{\sf V}\equiv V^{H\dag}_L V^{\ell}_L,\quad 
\widehat{\sf W}\equiv  V^{\widetilde{\ell}^c_{+}\dag}_L V^{H}_L.
\label{mixingmatrices}
}
One could argue that another mixing matrix $V^{\widetilde{\ell}^c_{+}\dag}_L V^{\ell}_L$ is also needed, but this is not the case because there is no gauge or Yukawa coupling between the SM doublet $l_L$ and the new fields. On the other hand, the matrices ${\sf W}\equiv V^{\widetilde{\ell}^c_{-}\dag}_L V_L^H$ and ${\sf Z}\equiv (V^\chi_R)^\dagger V^H_R$ introduced in \cite{delaguilaLeptonFlavorChanging2017,delaguilaInverseSeesawNeutrino2019} are both the identity in our model, since now the T-odd combination of the mirror-partner leptons and the T-even combination of the $\chi$ receive their masses from the same Yukawa Lagrangian $\Lag_{\widehat{Y}_H}$.

This completes the derivation of the full Lagrangian. Next we proceed to the calculation of the Goldstone potential, which is generated radiatively.

\subsection{New fermion contribution to LFV Higgs decays}

In the original LHT model one can define two different implementations of T-parity on the fermion fields. Depending on the T-parity realization, the singlet $\chi_R$ inside the SO(5) quintuplet $\Psi_R$ in eq.~(\ref{psiR}) can be chosen T-odd $(\chi_{\mathunderscore})_R$ or T-even $(\chi_{+})_R$. As we have shown, the T-odd case is incompatible with gauge invariance, resulting in an infinite contribution to lepton flavor violating (LFV) Higgs decays at order $v^2/f^2$ \cite{delaguilaInverseSeesawNeutrino2019}. However, the T-even option gives a finite result. In short, this is because the infinite contribution of mirror and mirror partner leptons cancels each other \cite{delaguilaLeptonFlavorChanging2017} but the individual contribution of a T-odd singlet is divergent. On the other hand the contribution of a T-even singlet is finite, as will be presented elsewhere.

As already emphasized, gauge invariance requires the singlet in the SO(5) quintuplet to be  
T-even. Our model verifies this requirement. Our $(\chi_+)_R$ has the same couplings with the SM charged leptons as the T-even singlet of the original LHT, so its contribution to LFV Higgs decays is finite as well. Moreover, to provide gauge invariant mass terms to all the fermions in our model, we had to enlarge the fermion field content. Firstly we have completed the left-handed SU(5) quintuplets in eq.~(\ref{newsu5multiplets}). Their combination with well defined T-parity now includes two new singlets $(\chi_{\pm})_L$ and two new doublets of mirror partners leptons $(\widetilde{l}^c_{\pm})_L$ apart from the usual doublets $l_L$ and $l_{HL}$ of SM and mirror leptons, respectively. And secondly we have introduced the right-handed quintuplet $\widehat{\Psi}_R$ in eq.~(\ref{newmultiplet}) including a T-even doublet of mirror partners $(\widetilde{l}^c_{+})_R$ and a T-odd singlet $(\chi_{\mathunderscore})_R$. One may argue that these new fields could reintroduce unwanted divergences in LFV Higgs decays, since in particular there is a new T-odd $(\chi_{\mathunderscore})_R$. To prove that this is actually not the case, below we will analyze their couplings to the SM charged leptons. They are needed to compute the divergences of the different classes of one-loop diagrams (see e.g. \cite{delaguilaLeptonFlavorChanging2017}). The relevant vertices come from the Lagrangians $\mathcal{L}_{F_L}$, $\mathcal{L}_{Y_H}$, $\mathcal{L}_{\widehat{Y}_H}$ and $\mathcal{L}_{Y}$ in eqs.~(\ref{LagFL}), (\ref{lagkappaa}), (\ref{Lagkappahat}) and (\ref{laglambda}), respectively.

The couplings between gauge bosons and left-handed fermions in $\mathcal{L}_{F_L}$ involve the gauge generators in eqs.~(\ref{generators1}) and (\ref{generators2}), that cannot connect the upper and lower components of the quintuplets, hence forbidding any coupling between the SM charged leptons and the new left-handed fermion fields. Concerning the Yukawa Lagrangian $\mathcal{L}_{Y_H}$, the new left-handed fields share quintuplets with $l_{1L}$ and $l_{2L}$ preventing any coupling to the SM charged leptons. On the other hand, the new Yukawa Lagrangian $\mathcal{L}_{\widehat{Y}_H}$ couples $l_{1L}$ and $l_{2L}$ to the multiplet $\widehat{\Psi}_R$ in eq.~(\ref{newmultiplet}) through the non linear sigma field $\widehat{\xi}$. Since it is built as the exponential of the new Goldstone bosons multiplying the broken combination of gauge generators, the same argument applied in $\mathcal{L}_{F_L}$ is also valid here. In particular, the T-odd right-handed singlet only couples to its T-even or T-odd left-handed counterpart through the new Goldstone fields. Therefore, its interactions are completely different to those of the original model with the T-odd singlet option. Finally, the Yukawa Lagrangian  $\mathcal{L}_{Y}$ only couples the right-handed SM charged leptons $\ell_R$ to $l_{1L}$ and $l_{2L}$ because the multiplets $\Psi_1^X$ and $\Psi_2^{X^*}$ in eq.~(\ref{PsiX}) are incomplete. As a consequence, in our model there are no vertices between the SM charged leptons and the new fermion fields.

Finally, the new neutral fields might still generate a divergent contribution through mixing with other fields that have direct couplings, as $\nu_{HR}$ for instance. Restricting ourselves to order $v^2/f^2$ the possible terms are presented in Table~\ref{table:mixing}. Among the one-loop diagrams listed in \cite{delaguilaLeptonFlavorChanging2017} one can distinguish two different cases. In the case where the diagram involves a Higgs coupling to two neutral fermions, that can be inferred from eq~(\ref{neutralmassmatrices}) and Table \ref{table:mixing} by substituting the \vev\ by a Higgs boson, at least a mass or one mixing insertion is required. On the other hand, one may check that before the mixing insertion the degree of divergence of this kind of topologies is at most of order $\log\Lambda$ in the cut-off regularization scheme. Therefore, the contribution of the new fermion fields to this kind of topologies is finite. This is because the mass contributes as $M/p^2$ to the loop and each mixing insertion is followed by the introduction of a new propagator contributing at least as $1/\slashed{p}$, with $p$ the corresponding momentum. In the other case, at least two mixing insertions are required. Furthermore the degree of divergence of the involved topologies is at most of order $\Lambda$ before the mixing insertions. Thus the contribution of the new fermion fields to these topologies is also finite. As a consequence, LFV Higgs decays are finite. This stems from two reasons: our new model is gauge invariant and we have chosen the appropriate fermion field representations.


\section{The Coleman-Weinberg potential}\label{BFM}

In this section we calculate the Coleman-Weinberg potential \cite{colemanRadiativeCorrectionsOrigin1973} generated by integrating out fermions and gauge bosons at one loop. For this purpose we apply the background field method (BFM) with a proper time cut-off, that allows the classification of divergences into quadratic and logarithmic. This is relevant because only logarithmic divergences in the Higgs mass are admissible. First we will derive generic expressions for the potential that then will be applied to our model in order to derive  the physical scalar masses and the Higgs potential.

\subsection{Integrating out fermions and gauge bosons in the background field method}

The BFM allows to find the divergent terms of a theory in a gauge invariant way, translating the divergences in the spacetime integration into those of the functional trace of a heat kernel in a new variable called proper time \cite{ballChiralGaugeTheory1989,neufeldOneloopFunctionalBerezinian1998}. This method has been extensively employed in the literature to study the renormalization of the linear realization of the SM \cite{dennerApplicationBackgroundFieldMethod1995,dittmaierDerivingNondecouplingEffects1995} and more recently has been applied to its non linear realization in \cite{buchallaCompleteOneLoopRenormalization2018}, where a master formula was derived in the dimensional regularization scheme using super-heat kernel tools \cite{neufeldSuperHeatKernelExpansionRenormalization1999}. The BFM is also useful when dealing with the Standard Model Effective Field Theory (SMEFT; see \cite{brivioStandardModelEffective2018} for a recent review). In particular, in \cite{buchallaMasterFormulaOneloop2019} a master formula that includes the effects of bosonic operators up to dimension six is applied to the calculation of the Renormalization Group equations in the context of the SMEFT. 

Our starting point is a general Lorentz-invariant four-dimensional action containing real bosonic fields and operators at most bilinear in the fermion fields,
\eq{
S\left[\varphi^i,A_{\mu}^a,\psi^{b},\overline{\psi}^{b}\right]=\int d^4x\,\Lag\left(\varphi^i,A_{\mu}^a,\psi^{b},\overline{\psi}^{b}\right),
}
where Latin indices refer to the different species of bosons and fermions in our theory.\footnote{If the bosonic fields are complex they are split into real and imaginary parts.} To obtain the generating functional of Green functions, we couple the fields to external sources
\eq{\label{Z}
Z[j_i,&J_{a}^{\mu},\rho_{b},\overline{\rho}_{b}]=\int\left[D\varphi D A_{\mu} D\psi D\overline{\psi}\right]\nn\\&\times\exp\left\{i\left(S+ \left\langle j_i\varphi^i+J_{a}^{\mu}A^a_{\mu}+\overline{\psi}^{b}\rho_{b}+\overline{\rho}_{b}\psi^{b}\right\rangle\right)\right\},
}
where $Z=e^{iW}$ with $W$ the generating functional of connected Green functions and $\langle\cdots\rangle$ stands for integration over spacetime. The path integral is normalized to $Z\left[0\right]=1$. The classical, or background, fields are the solutions of the classical equations of motion (EoM)
\eq{
\left.\frac{\delta S}{\delta \varphi^i}\right|_{\varphi_{\textrm{cl}}^i}\!\!\!+j_i&=0, \quad \left.\frac{\delta S}{\delta A_{\mu}^a}\right|_{A^a_{\mu,\textrm{cl}}}\!\!\!+J^{\mu}_a=0,\quad \\
\frac{\delta S}{\delta \overline{\psi}^{b}}\biggm|_{\psi_{\textrm{cl}}^{b}}+\rho_{b}&=0,\quad\frac{\delta S}{\delta \psi^{b}}\biggm|_{\overline{\psi}^{b}_{\textrm{cl}}}-\overline{\rho}_{b}=0.
}
(`cl' stands for classical and the Grassmannian character of fermion fields has been used.) 

In order to integrate out gauge bosons and fermions and get the scalar potential, we perform a change of variables in the path integral consisting in a linear split of each gauge or fermion field in two parts: the background field and the (quantum) fluctuating field which will be the new variable of integration in the path integral,
\eq{\label{split}
A_{\mu}^{a}&=A^{a}_{\mu,\textrm{cl}}+\textrm{a}_\mu^{a},\\
\psi^{b}&=\psi^{b}_{\textrm{cl}}+\chi^{b},\\
\overline{\psi}^{b}&=\overline{\psi}_{\textrm{cl}}^{b}+\overline{\chi}^{b}.
}
The scalar fields are fixed by the EoM, $\varphi^i=\varphi_{\textrm{cl}}^i$.
The background fields only appear as external legs in the Feynman diagrams whereas the fluctuating fields only occur in loops. Substituting in the Lagrangian and keeping terms up to second order in fermion and gauge fluctuations one may parametrize the action as
\eq{
S&+\left\langle j_i\varphi^i+J_a^{\mu}A^a_{\mu}+\overline{\psi}^{b}\rho_{b}+\overline{\rho}_{b}\psi^{b}\right\rangle \nn\\&= S^{(0)}+\left\langle j_i\varphi^i_{\textrm{cl}}+J^{\mu}_a A^{a}_{\mu,\textrm{cl}}+\overline{\psi}^{b}_{\textrm{cl}}\rho_{b}+\overline{\rho}_{b}\psi^{b}_{\textrm{cl}}\right\rangle\nn\\&
+S^{(2)}\left[\varphi^i_{\textrm{cl}};\textrm{a}_{\mu}^a,\chi^{b},\overline{\chi}^{b}\right].
}
The first term on the r.h.s is the classical action evaluated in the background fields and the last is quadratic in the fluctuations,\footnote{The part of the action that is linear in the fluctuating fields is identically zero since it is proportional to the EoM and the background fields are on shell.} and takes the general form 
\eq{
S^{(2)}\left[\varphi^i_{\textrm{cl}};\textrm{a}_{\mu}^a,\chi^{b},\overline{\chi}^{b}\right]=\int d^4 x\, \Lag^{\left(2\right)}\left(\varphi^i_{\textrm{cl}};\textrm{a}^a_{\mu},\chi^{b},\overline{\chi}^{b}\right),
} 
with
\eq{\label{actionparametrization}
\Lag^{\left(2\right)}\left(\varphi_{\textrm{cl}}^i;\textrm{a}^a_{\mu},\chi^{b},\overline{\chi}^{b}\right)&=-\frac{1}{2}\phi^k A^{\dis l}_{k}\phi_l+\overline{\chi}^{b}\left(i\slashed{\partial}-G\right)_{b c}\chi^{c}\nn\\&\equiv-\frac{1}{2}\phi^T A \phi+\overline{\chi}B\chi,
}
where $\phi^i$ collects the bosonic (gauge) fluctuations, and all the dependence on the background fields $\varphi^i_{\textrm{cl}}$ is encoded in the matrices $A$ and $G$, with $\{k,l\}$ indices of any type. After integrating by parts, the interaction between bosonic fields is given by second order differential operators 
\eq{\label{Adef}
A=D^{\mu}D_{\mu}+V, \quad \textrm{with} \quad D_{\mu}=\partial_{\mu}+N_{\mu},
}
where $N_{\mu}$ and $V$ depend on the background scalar fields, and the most general Lorentz structure of the scalar-fermion interaction is
\eq{\label{Gdef}
G = r P_R + l P_L .
}
To express the bosonic interactions in the canonical form of eq.~(\ref{Adef}), one must introduce appropriate gauge-fixing terms for the fluctuating gauge fields, still preserving the gauge invariance of the one-loop effective action \cite{Abbott:1981ke}.
Now we redefine all the fluctuating gauge fields to have the same sign in the derivatives of time and space components,
\eq{\label{redef1}
\widetilde{\textrm{a}}^a_{\mu}=\left(i \textrm{a}^a_0, \textrm{a}_i^a\right)\equiv M_\mu^{\;\nu}\textrm{a}^a_{\nu},\quad M_{\mu}^{\;\nu}=\textrm{diag}\left(i,1,1,1\right).
} 
The contravariant vector $\textrm{a}^{a\mu}=\left(\textrm{a}^a_0,-\textrm{a}_i^a\right)$ transforms with the inverse matrix 
\eq{\label{redef2}
\widetilde{\textrm{a}}^{a\mu}&=\left(M^{-1}\right)^{\mu}_{\;\nu} \textrm{a}^{a\nu}=\left(-i \textrm{a}^a_0,-\textrm{a}^a_i\right),\quad \\
\left(M^{-1}\right)^{\mu}_{\;\nu}&=\textrm{diag}\left(-i,1,1,1\right),
}
implying that $\widetilde{\textrm{a}}^a_{\mu}=-\widetilde{\textrm{a}}^{a\mu}$.\footnote{The auxiliary matrix $M$ appears in intermediate steps of the calculation but it will cancel at the end.} In terms of these new gauge fields, the kinetic term reads
\eq{
\Lag_{\textrm{kin,g}}=-\frac{1}{2}\widetilde{\textrm{a}}^a_{0}\partial^2 \widetilde{\textrm{a}}^{a}_0-\frac{1}{2}\widetilde{\textrm{a}}^a_{i}\partial^2 \widetilde{\textrm{a}}^{a}_i=-\frac{1}{2}\left(-\widetilde{\textrm{a}}^{a\mu}\right)\delta^{\nu}_{\mu}\partial^2\widetilde{\textrm{a}}^a_{\nu}.
}
Comparing with eq.~(\ref{actionparametrization}) we have $\phi^k=-\widetilde{\textrm{a}}^{a\mu}=\widetilde{\textrm{a}}^a_{\mu}$ and $\phi_l=\widetilde{\textrm{a}}^{a}_{\mu}$. The advantage of this redefinition of the fluctuating gauge fields is that the functions at both sides of the operator $A$ are the same, which is necessary to later perform a Gaussian integration. 

The expansion of the action to second order in the fluctuations is enough to get the generating functional $W$ to one loop,
\eq{
W=W_{L=0}+W_{L=1}+\textrm{higher order corrections,}
}
where $L$ is the number of loops. Then, to this order, the generating functional $Z$ can be written as $Z=Z_{L=0}Z_{L=1}$. The first factor is
\eq{
Z_{L=0}=e^{iW_{L=0}}=e^{i\left(S^{(0)}+\left\langle j_i\varphi^i_{\textrm{cl}}+J^{\mu}_{a}A^{a}_{\mu,\textrm{cl}}+\overline{\psi}^b_{\textrm{cl}}\rho_b+\overline{\rho}_b\psi^b_{\textrm{cl}}\right\rangle \right)},
}
a constant for the path integral, independent of the quantum variables. Taking logarithms,
\eq{
W_{L=0}=S^{(0)}+\left\langle j_i\varphi^i_{\textrm{cl}}+J^{\mu}_a A^a_{\mu,\textrm{cl}}+\overline{\psi}^b_{\textrm{cl}}\rho_b+\overline{\rho}_b\psi^b_{\textrm{cl}}\right\rangle ,
}
which after subtracting the source term is nothing  but the classical action. Focusing on the contribution of gauge bosons and fermions to the scalar potential, the one loop correction to the generating functional $W$ comes from the quadratic Lagrangian in eq.~(\ref{actionparametrization}), 
\eq{
e^{iW_{L=1}}&=\int\left[D\phi D\chi D\overline{\chi} \right]e^{iS^{\left(2\right)}} \nn\\&=\int\left[D\phi D\chi D\overline{\chi} \right]e^{i\int d^4x\dis\left(-\frac{1}{2}\phi^T A \phi+\overline{\chi}B\chi\right)}.
}
Performing the Gaussian integration over the gauge and fermion fields in the path integral and taking logarithms, the one-loop generating functional reads
\eq{\label{detAB}
W_{L=1}=\frac{i}{2}\log\textrm{Det}\,A-i\log\textrm{Det}\,B
}
where Det stands for the functional determinant and capital letters indicate that it involves a spacetime integration. Since it does not depend on the source term, $W_{L=1}$ can be interpreted as the one-loop effective action. Squaring the operator $B$ as in {\cite{ballChiralGaugeTheory1989,BerredoPeixoto2001ANO}}, we may write
\eq{
\log\textrm{Det}\,B = \frac{1}{2}\log\textrm{Det}\left(B B^c\right),
}
where $B^c=-i\slashed{\partial}-G^{\dagger}$. Then
\eq{\label{delta}
W_{L=1}
=\frac{i}{2}\log \textrm{Sdet}\left(\begin{array}{cc}
A & 0\\
0 & B B^c
\end{array}\right)\equiv\frac{i}{2}\log \textrm{Sdet}\dis\Delta.
}
In this compact notation borrowed from supersymmetry, $\textrm{Sdet}M$ stands for the superdeterminant or Berezinian of a supermatrix $M$,
\eq{
M=\left(\begin{array}{cc}
a & \alpha\\
\beta & b
\end{array}\right),\quad \textrm{Sdet}\,M \equiv \textrm{Det}\left(a-\alpha b^{-1} \beta\right)/\textrm{Det}\,b,
}
where the entries $a,b$ ($\alpha,\beta$) are bosonic (fermionic) variables. In our case $\alpha=\beta=0$. Notice that $\Delta$ can be written in the canonical form,
\eq{\label{canonical}
\Delta=\left(\partial^{\mu}+\Lambda^{\mu}\right)\left(\partial_{\mu}+\Lambda_{\mu}\right)+Y,
}
expression which holds in our case because our starting Lagrangian is, at most, bilinear in the fermion fields \cite{buchallaCompleteOneLoopRenormalization2018,buchallaMasterFormulaOneloop2019}.
Going to the Euclidean spacetime, performing the usual change of variables in the time coordinate, $t=-i\tau$, we can rewrite the one-loop effective action
\eq{
W_{L=1}^E=-iW_{L=1}=\frac{1}{2}\log\textrm{Det}\left(-\Delta^E\right)=\frac{1}{2}\textrm{Str}_{E}\log\left(-\Delta^E\right),
}
where $\textrm{Str}_{E}$ is the supertrace ($\text{Str}M=\textrm{tr}\,a-\textrm{tr}\,b$) that also includes a Euclidean spacetime integration.\footnote{The supertrace and the superdeterminant have the same properties as the usual trace and determinant with respect to the logarithms.} In the last expression, the operator $\Delta^E$ is defined as 
\eq{
\Delta^E=-\Delta=\left(\partial^{E}_{\mu}+\Lambda^{E}_{\mu}\right)\left(\partial^E_{\mu}+\Lambda^E_{\mu}\right)-Y^E,
}
where the Euclidean versions of the matrices above verify
\eq{\label{minktoeuc}
\Lambda_{\mu}\left(t,\vec{x}\right)&=\begin{cases}\Lambda_{0}\left(t,\vec{x}\right)=i\Lambda_{0}^{E}\left(\tau,\vec{x}\right)\\\Lambda_{i}\left(t,\vec{x}\right)=\Lambda_{i}^{E}\left(\tau,\vec{x}\right),\end{cases}\\
Y\left(t,\vec{x}\right)&=Y^E\left(\tau,\vec{x}\right).
}
To obtain the divergences of the one-loop functional, we rewrite it using the proper time or Schwinger representation \cite{ivanovHeatKernelProper2020}:
\eq{
W_{L=1}^E&=-\frac{1}{2}\int^{\infty}_0\frac{ds}{s}\textrm{Str}_{E}\dis e^{s\Delta^E} \nn\\&=-\frac{1}{2}\int^{\infty}_0\frac{ds}{s}\int d^4x^E\dis\textrm{str}\
{\bf K}(s;x^E;x^E),
}
where the integrand is the supertrace of the heat kernel of the elliptic operator $\Delta^E$. The reason to go to Euclidean spacetime is that the heat kernel is better behaved \cite{vassilevichHeatKernelExpansion2003}.
DeWitt \cite{DeWitt:1965jb} proposes the following ansatz for the heat kernel in the limit $s\rightarrow 0$,
\eq{\label{deWitt}
{\bf K}(s;x^E;x^E)
=\frac{1}{\left(4\pi s\right)^{2}}\sum_{n=0}^{\infty}a_{n}\left(x^E,x^E\right)s^n.
}
The coefficients $a_{n}$ are the so called Seeley-DeWitt coefficients \cite{DeWitt:1965jb,Seeley:1967ea} which are completely regular. As we will show below, the divergences we are looking for are proportional to the first coefficients. The integral converges in the upper limit.

To regularize the integral in the lower limit, let us focus in the units of the proper time variable $s$. Since the argument of the exponential is dimensionless, $s$ has units of inverse mass squared. Using a proper time cut-off in the lower limit we may write \cite{tomsQuadraticDivergencesQuantum2011}
\eq{
W_{L=1,\textrm{reg}}^E=-\frac{1}{2}\int^{\infty}_{\Lambda^{-2}}\frac{ds}{s}\int d^4x^E\dis\textrm{str}\dis {\bf K}(s;x^E;x^E),
}
with $\Lambda\rightarrow \infty$ the usual energy cut-off. Then we can use the DeWitt expansion in eq.~(\ref{deWitt}) to solve the integral in the proper time variable to find the divergences in the lower limit:
\eq{
\label{DWdivergences}
\int^{E_0^{-2}}_{\Lambda^{-2}}\frac{ds}{s}s^{n-2}\sim \begin{cases}
\frac{1}{2-n}\Lambda^{4-2n} & n<2\\
\log\Lambda^2 & n=2\\
-\frac{1}{n-2}
\Lambda^{2n-4} & n>2
\end{cases}
}
where we have introduced an upper limit $E_0^{-2}$ related to the maximum value of the proper time variable for which the DeWitt expansion of the heat kernel is valid. Actually this energy scale should appear in the argument of $\log\Lambda^2$ making it dimensionless, but it is omitted here and in the following. As previously advertised, the divergences are found just in the first three coefficients.\footnote{The coefficient $a_0=1$ is independent of the background fields, being only important if we were dealing with gravity, since it contributes to the vacuum energy.} Since we are interested in quadratic and logarithmic divergences, only the expressions of $a_1$ and $a_2$ are needed. These can be found in \cite{ballChiralGaugeTheory1989} and their values in Euclidean spacetime are
\eq{
a_1\left(x^E,x^E\right)&=-Y^E\\
a_2\left(x^E,x^E\right)&=\frac{1}{12}Z^E_{\mu\nu}Z^E_{\mu\nu}+\frac{1}{2}Y^{E 2},
\quad \\
Z^E_{\mu\nu}&=\partial^E_{\mu}\Lambda^E_{\nu}-\partial^E_{\nu}\Lambda^E_{\mu}+\left[\Lambda^E_{\mu},\Lambda^E_{\nu}\right].
}
Writing everything together, the effective action reads
\eq{
W_{L=1,\textrm{reg}}^E=&-\frac{1}{2}\frac{1}{16\pi^2}\int d^4x^E \textrm{str}\dis \Big[-Y^E\Lambda^2\nn\\&+\Big(\frac{1}{12}Z^E_{\mu\nu}Z^E_{\mu\nu}+\frac{1}{2}Y^{E 2}\Big)\log\Lambda^2\Big].
}
Turning back to the Minkowski spacetime using eq.~(\ref{minktoeuc}), the divergent part of the Lagrangian at one loop is
\eq{\label{1looplag}
\Lag^{\textrm{div}}_{L=1}=-\frac{1}{32\pi^2}  \textrm{str}\dis \left[Y\Lambda^2-\left(\frac{1}{12}Z^{\mu\nu}Z_{\mu\nu}+\frac{1}{2}Y^{2}\right)\log\Lambda^2\right].
}
In order to find the scalar potential we can ignore all the interactions involving derivatives of the background scalar fields. Then only $Y$ and the commutator $[\Lambda^\mu,\Lambda^\nu]$ in $Z^{\mu\nu}$ are needed. Besides, in our model $N^\mu=0$. Reversing the sign in previous expression we obtain
\eq{
\mathcal{V}^{\textrm{div}}_{L=1}\subset -\Lag^{\textrm{div}}_{L=1}.
}
Finally, using equations (\ref{delta}) and (\ref{canonical}), the expressions for the matrices $\Lambda_{\mu}$ and $Y$ read
\eq{
\Lambda^{\mu}=\left(\begin{array}{cc}
N^{\mu} & 0\\
0 & \frac{i}{2}\left(G\gamma^{\mu}-\gamma^{\mu}G^{\dagger}\right)
\end{array}\right),
}
and
\eq{
Y=\left(\begin{array}{cc}
V & 0\\
0 & Y_{22}
\end{array}\right),
}
with
\eq{
Y_{22}=&-\frac{i}{2}\partial_{\mu}\left(G\gamma^{\mu}-\gamma^{\mu}G^{\dagger}\right) \nn\\&+\frac{1}{4}\left(G\gamma^{\mu}G\gamma_{\mu}-\gamma^{\mu}G^{\dagger}G\gamma_{\mu}+\gamma^{\mu}G^{\dagger}\gamma_{\mu}G^{\dagger}\right).
}

Evaluating the supertrace of the matrix $Y$ ignoring all the derivative interactions involving background scalars and performing the traces in spinor space \cite{patelPackageXMathematicaPackage2015} using the definition of $G$, the quadratic and logarithmic parts of the scalar potential are given by
\eq{\label{quadraticdiv}
\mathcal{V}_{L=1}^{\Lambda^2}=\frac{\Lambda^2}{32\pi^2}\textrm{tr}\dis V-\frac{\Lambda^2}{8\pi^2}\textrm{tr}\left(l r\right),
}
and
\eq{\label{logdiv}
\mathcal{V}_{L=1}^{\log\Lambda^2}=-\frac{1}{64\pi^2}\log\Lambda^2\textrm{tr}\dis V^2+\frac{1}{16\pi^2}\log\Lambda^2\textrm{tr}\left(l r l r\right).
}
These are the two master formulas that we will employ in the following.

\subsection{Quadratic and logarithmic corrections to the scalar potential}

To obtain the quadratic and logarithmic contributions to the scalar one loop effective potential, we have to evaluate the expression for the matrices $V$, $l$ and $r$ defined in eqs.~(\ref{actionparametrization}), (\ref{Adef}), (\ref{Gdef}). In this work we neglect the contribution of the lightest SM lepton Yukawa interactions in eqs.~(\ref{laglambda}) as well as those of the lightest SM quarks, since their Yukawa couplings are parametrically small. Thus, the leading contributions come from the interaction between scalars and gauge bosons in eqs.~(\ref{lagS}), (\ref{lagShat}), the Yukawa Lagrangian for the heavy leptons in eqs.~(\ref{lagkappaa}) and (\ref{Lagkappahat}) with the field content of eqs.~(\ref{PsiRTeven}), (\ref{newsu5multiplets}) and (\ref{newmultiplet}), plus the analogous heavy quark Yukawa Lagrangians given by
\eq{\label{Lagkappaq}
\Lag_{Y_{q_H}}=-\kappa_q f\left(\overline{\Psi}^q_2\xi+\overline{\Psi}^q_1\Sigma_0\xi^{\dagger}\right)\Psi^q_R+\hc
}
and
\eq{
\Lag_{\widehat{Y}_{q_H}}=-\widehat{\kappa}_q f\left(\overline{\Psi}^q_2\hat{\xi}-\overline{\Psi}^q_1\Sigma_0\hat{\xi}^{\dag}\right)\widehat{\Psi}^q_R+\hc,\label{Lagkappahatq}
}
where $\kappa_q$ and $\widehat{\kappa}_q$ are matrices in flavor space. Finally, one needs the top Yukawa Lagrangian \cite{Han:2005ru,hubiszPhenomenologyLittlestHiggs2005} which implements the collective symmetry breaking to avoid quadratic divergences on the Higgs boson mass proportional to the top Yukawa,
\eq{\label{toplag}
\mathcal{L}_t&=-i\frac{\lambda_1}{4}f\epsilon_{ijk}\epsilon_{xy}\left[\left(\overline{Q}_1\right)_i\Sigma_{jx}\Sigma_{ky}+\left(\overline{Q}_2\Sigma_0\Omega\right)_i\widetilde{\Sigma}_{jx}\widetilde{\Sigma}_{ky}\right]t_R\nn\\&-\frac{\lambda_2}{\sqrt{2}} f\left(\overline{T}'_{1L} T'_{1R}+\overline{T}'_{2L} T'_{2R}\right)+\hc
}
with $\{i,j,k\}=1,2,3$ and $\{x,y\}=4,5$.
The quark multiplets in eqs.~(\ref{Lagkappaq}), (\ref{Lagkappahatq}) are defined by
\eq{\label{quarkmultiplets}
\Psi^q_{1}&=\left(\begin{array}{c}
-i\sigma^{2}q_{1L}\\
i\chi^q_{1L}\\
-i\sigma^2\widetilde{q}^c_{1L}
\end{array}\right),\quad \Psi^q_{2}=\left(\begin{array}{c}
-i\sigma^2\widetilde{q}^c_{2L}\\
i\chi^q_{2L}\\
-i\sigma^{2}q_{2L}
\end{array}\right), \quad \\ 
\Psi^q_{R}&=\left(\begin{array}{c}
-i\sigma^{2}(\widetilde{q}^{c}_{\mathunderscore})_{R}\\
i\left(\chi^q_{+}\right)_{R}\\
-i\sigma^{2}q_{HR}
\end{array}\right), \quad 
\widehat{\Psi}^q_{R}=\left(\begin{array}{c}
-i\sigma^{2}(\widetilde{q}^{c}_{+})_{R}\\
i\left(\chi^q_{\mathunderscore}\right)_{R}\\
0_2
\end{array}\right)
}
where
\eq{
q_{rL}=\left(\begin{array}{c}
u_{rL}\\
d_{rL}
\end{array}\right), \quad \widetilde{q}^c_{rL}=\left(\begin{array}{c}
\widetilde{u}_{rL}\\
\widetilde{d}_{rL}
\end{array}\right), \quad r=1,2
}
and
\eq{
q_{HR}=\left(\begin{array}{c}
u_{HR}\\
d_{HR}
\end{array}\right), \quad \left(\widetilde{q}^c_{\pm}\right)_{R}=\left(\begin{array}{c}
\left(\widetilde{u}^c_{\pm}\right)_R\\
\left(\widetilde{d}^c_{\pm}\right)_R
\end{array}\right)
}
with analogous transformation properties under the gauged subgroup and T-parity as those for leptons (see eqs.~(\ref{PsiRtransformation}), (\ref{TevenPsiRtransformation}), (\ref{su5multipletstransformation}), (\ref{newmultiplet})). The multiplets that appear in eq.~(\ref{toplag}) are given by
\eq{\label{quarktopmultiplets}
Q_{1}=\left(\begin{array}{c}
-i\sigma^{2}\mathcal{T}_{1L}\\
i\dis T'_{1L}\\
0_2
\end{array}\right),\quad Q_{2}=\left(\begin{array}{c}
0_2\\
i\dis T'_{2L}\\
-i\sigma^{2}\mathcal{T}_{2L}
\end{array}\right),
}
where
\eq{
\mathcal{T}_{rL}=\left(\begin{array}{c}
t_{rL}\\
b_{rL}
\end{array}\right), \quad r=1,2
}
with the transformation properties
\eq{
Q_1 \xrightarrow{G_g} V^{*}_g Q_1,\quad 
Q_2 \xrightarrow{G_g} V_g Q_2,\quad  Q_1 \xrightarrow{\textrm{T}} \Omega\Sigma_0 Q_2
}
and the $\su2$ singlets
\eq{
T'_{1R}\xrightarrow{\textrm{T}} T'_{2R}.
}
In what follows, the Higgs potential is parametrized as
\eq{
\label{Hpot}
\mathcal{V}_{\textrm{Higgs}}=\mu^2 \left(H^{\dagger}H\right)+\lambda \left(H^{\dagger}H\right)^2.
}

\subsubsection{Gauge boson contribution to the scalar potential}

Using eq.~(\ref{split}) for the fluctuating gauge bosons in eqs.~(\ref{lagS}) and (\ref{lagShat}),
\eq{
\mathcal{L}^{(2)}_{S}+
\mathcal{L}^{(2)}_{\widehat{S}}
\supset \frac{f^2}{2}\sum_{j,k=1}^2 &\Big[g^2\omega^a_{j\mu}\omega^{b\mu}_k\textrm{tr}\left(Q^a_jQ^b_k+Q^a_j\Sigma Q^{b T}_k\Sigma^{\dagger}\right)\nn\\
&+g'^2 b_{j\mu}b^{\mu}_k\textrm{tr}\left(Y_j Y_k+Y_j\Sigma Y^{T}_k\Sigma^{\dagger}\right)\nn\\
&-g g' \omega^a_{j\mu}b^{\mu}_k\textrm{tr}\left(Q^a_j\Sigma Y^{T}_k\Sigma^{\dagger}\right) \nn\\
&-g g' b^{\mu}_j\omega^b_{k\mu}\textrm{tr}\left(Q^b_k\Sigma Y^{T}_j\Sigma^{\dagger}\right)\Big]\nn\\
+\frac{f^2}{2}\sum_{j,k=1}^2
&\Big[g^2\omega^a_{j\mu}\omega^{b\mu}_k\textrm{tr}\left(Q^a_jQ^b_k+Q^a_j\widehat{\Sigma} Q^{b T}_k\widehat{\Sigma}^{\dagger}\right)\nn\\
&+g'^2 b_{j\mu}b^{\mu}_k\textrm{tr}\left(Y_j Y_k+Y_j\widehat{\Sigma} Y^{T}_k\widehat{\Sigma}^{\dagger}\right)\nn\\
&-g g' \omega^a_{j\mu}b^{\mu}_k\textrm{tr}\left(Q^a_j\widehat{\Sigma} Y^{T}_k\widehat{\Sigma}^{\dagger}\right)\nn\\
&-g g' b^{\mu}_j\omega^b_{k\mu}\textrm{tr}\left(Q^b_k\widehat{\Sigma} Y^{T}_j\widehat{\Sigma}^{\dagger}\right)\Big],
}
where the background scalars are parametrized in the non linear sigma fields $\Sigma$, $\widehat{\Sigma}$ and $\omega^a_{j\mu}$, $b_{j\mu}$ are the fluctuations of the gauge bosons associated to \su2 and \u1 respectively. As already mentioned, using the appropriate gauge for these fluctuating fields, their kinetic terms take the canonical form. The redefinition of the gauge fields as in eqs.~(\ref{redef1}) and (\ref{redef2}),
\eq{
\omega^a_{j\mu}\omega^{b\mu}_k&=-\left(-\tilde{\omega}^{a\mu}_j\right)\delta^{\nu}_{\mu}\tilde{\omega}^b_{k\nu},\\
b_{j\mu}b^{\mu}_k&=-\left(-\tilde{b}^{\mu}_j\right)\delta^{\nu}_{\mu}\tilde{b}_{k\nu},\\
\omega^a_{j\mu}b^{\mu}_k &= -\left(-\tilde{\omega}^{a\mu}_j\right)\delta^{\nu}_{\mu}\tilde{b}_{k\nu},
}
allows us to find the corresponding matrix $V$
\eq{\label{V}
V&=\left(\begin{array}{cc} V_{11} & V_{12} \\ V_{21} & V_{22}
\end{array}\right),
}
where
\eq{
V_{11}&=f^2 \delta_{\mu}^{\nu}
g^{2}\textrm{tr}\left(2Q_{j}^{a}Q_{k}^{b}+Q_{j}^{a}\Sigma Q_{k}^{bT}\Sigma^{\dagger}+Q_{j}^{a}\widehat{\Sigma}Q_{k}^{bT}\widehat{\Sigma}^{\dagger}\right)
\\
V_{12}&=-f^2 \delta_{\mu}^{\nu}
gg'\textrm{tr}\left(Q_{j}^{a}\Sigma Y_{k}^{T}\Sigma^{\dagger}+Q_{j}^{a}\widehat{\Sigma}Y_{k}^{T}\widehat{\Sigma}^{\dagger}\right)
\\
V_{21}&=-f^2 \delta_{\mu}^{\nu}
gg'\textrm{tr}\left(Q_{k}^{b}\Sigma Y_{j}^{T}\Sigma^{\dagger}+Q_{k}^{b}\widehat{\Sigma}Y_{j}^{T}\widehat{\Sigma}^{\dagger}\right)
\\
V_{22}&=f^2 \delta_{\mu}^{\nu}
g'^{2}\textrm{tr}\left(2Y_{j}Y_{k}+Y_{j}\Sigma Y_{k}^{T}\Sigma^{\dagger}+Y_{j}\widehat{\Sigma}Y_{k}^{T}\widehat{\Sigma}^{\dagger}\right)
.
}
According to eq.~(\ref{quadraticdiv}), to obtain the quadratic divergences of this sector, we must take the trace of the matrix in eq.~(\ref{V}). $V$ is a block matrix with gauge group as well as Lorentz indices in each block. Taking the trace implies $a=b$, $j=k$, $\mu=\nu$ in each block and then sum the diagonal terms of the matrix (sum over repeated indices in their corresponding range is understood)
\eq{
\mathcal{V}^{\Lambda^2}_{L=1,g}=\frac{\Lambda^2}{8\pi^2} \Big[&g^{2}f^{2}\textrm{tr}\left(Q_{j}^{a}\Sigma Q_{j}^{aT}\Sigma^{\dagger}+Q_{j}^{a}\widehat{\Sigma}Q_{j}^{aT}\widehat{\Sigma}^{\dagger}\right) \nn\\
&+g'^{2}f^{2}\textrm{tr}\left(Y_{j}\Sigma Y_{j}^{T}\Sigma^{\dagger}+Y_{j}\widehat{\Sigma}Y_{j}^{T}\widehat{\Sigma}^{\dagger}\right)\Big]
}
where a global facor 4 comes from $\delta^{\mu}_{\mu}=4$ and irrelevant constant terms have been dropped. This part of the potential contains a mass term for the triplet $\Phi$ and a quartic Higgs coupling
\eq{\label{phimass1}
\mathcal{V}^{\Lambda^2}_{L=1,g}\supset& \frac{\Lambda^2}{4\pi^2}\left(g^2+g'^2\right)\textrm{tr}\left(\Phi^{\dagger}\Phi\right)
\nn\\&+\frac{1}{16\pi^2}\frac{\Lambda^2}{f^2}\left(g^2+g'^2\right)\left(H^{\dagger}H\right)^{2}
}
and no mass terms for the rest of the physical scalars. Since the leading order of the mass of the triplet $\Phi$ and the Higgs quartic coupling is $\Lambda^2\sim 16\pi^2 f^2$, we will neglect all the logarithmic contributions to these operators in the following. 

To evaluate the logarithmic divergences, from eq.~(\ref{logdiv}) we must take the trace of $V^2$. To construct this matrix we must pair the indices of both $V$ factors. For the sake of clarity, the left upper block $V^2_{11}$ would contain
\eq{
V^2_{11}\supset &f^2 \delta_{\mu}^{\nu}g^{2}\textrm{tr}\left(2Q_{j}^{a}Q_{k}^{b}+Q_{j}^{a}\Sigma Q_{k}^{bT}\Sigma^{\dagger}+Q_{j}^{a}\widehat{\Sigma}Q_{k}^{bT}\widehat{\Sigma}^{\dagger}\right)\nn\\
\times &f^2 \delta_{\lambda}^{\mu}g^{2}\textrm{tr}\left(2Q_{k}^{b}Q_{l}^{c}+Q_{k}^{b}\Sigma Q_{l}^{cT}\Sigma^{\dagger}+Q_{k}^{b}\widehat{\Sigma}Q_{l}^{cT}\widehat{\Sigma}^{\dagger}\right).
}
Then we repeat the same procedure applied above to take the trace. Again a global factor 4 will appear from the trace over the Lorentz indices $\delta^{\mu}_{\nu}\delta^{\nu}_{\mu}=\delta^{\mu}_{\mu}=4$ leading to
\eq{
\mathcal{V}_{L=1,g}^{\log\Lambda^2}&=-\frac{1}{16\pi^2}f^4\log\Lambda^2 \nn\\
&\times\Big[g^4 \textrm{tr}\left(2Q_{j}^{a}Q_{k}^{b}+Q_{j}^{a}\Sigma Q_{k}^{bT}\Sigma^{\dagger}+Q_{j}^{a}\widehat{\Sigma}Q_{k}^{bT}\widehat{\Sigma}^{\dagger}\right)\nn\\&\qquad\times\textrm{tr}\left(2Q_{k}^{b}Q_{j}^{a}+Q_{k}^{b}\Sigma Q_{j}^{aT}\Sigma^{\dagger}+Q_{k}^{b}\widehat{\Sigma}Q_{j}^{aT}\widehat{\Sigma}^{\dagger}\right)\nn\\
&\quad+ 2 g^2g'^2 \textrm{tr}\left(Q_{j}^{a}\Sigma Y_{k}^{T}\Sigma^{\dagger}+Q_{j}^{a}\widehat{\Sigma}Y_{k}^{T}\widehat{\Sigma}^{\dagger}\right) \nn\\&\qquad\times\textrm{tr}\left(Q_{j}^{a}\Sigma Y_{k}^{T}\Sigma^{\dagger}+Q_{j}^{a}\widehat{\Sigma}Y_{k}^{T}\widehat{\Sigma}^{\dagger}\right)\nn\\
&\quad+g'^4\textrm{tr}\left(2 Y_j Y_k + Y_{j}\Sigma Y_{k}^{T}\Sigma^{\dagger}+Y_{j}\widehat{\Sigma}Y_{k}^{T}\widehat{\Sigma}^{\dagger}\right)
\nn\\&\qquad\times\textrm{tr}\left(2 Y_k Y_j + Y_{k}\Sigma Y_{j}^{T}\Sigma^{\dagger}+Y_{k}\widehat{\Sigma}Y_{j}^{T}\widehat{\Sigma}^{\dagger}\right)\Big].
}
This part of the potential contains a contribution to the $\mu^2$ term in the Higgs potential and a mass term for the triplet $\widehat{\omega}$
\eq{\label{omegamass1}
\mathcal{V}_{L=1,g}^{\log\Lambda^2}&\supset  f^2\log\Lambda^2\left(\frac{3g^4}{8\pi^2}+\frac{g'^4}{40\pi^2}\right)\left(H^{\dagger}H\right)\nn\\&\quad+\frac{g^4}{\pi^2}f^2\log\Lambda^2\textrm{tr}\left(\widehat{\omega}\widehat{\omega}\right)\nn\\
&=\frac{1}{16\pi^2}\log\Lambda^2\left(3g^2 M^2_{W_H}+ g'^2M^2_{A_H}\right)\left(H^{\dagger}H\right)\nn\\&\quad+\frac{g^2}{2\pi^2}M^2_{W_H}\log\Lambda^2\textrm{tr}\left(\widehat{\omega}\widehat{\omega}\right),
}
where we have used $M^2_{W_H}=2 g^2 f^2$ and $M^2_{A_H}=\frac{2}{5}g'^2f^2$. These masses are naturally smaller than the mass of the triplet $\Phi$ since $\log\Lambda^2$ is parametrically of order one and there is a suppression of $16\pi^2$ which allows masses of order of the electroweak scale. 
No mass term is induced for the physical scalar $\widehat{\eta}$ by gauge interactions.

\subsubsection{Fermion contribution to the scalar potential}

In order to obtain the heavy leptons, heavy quarks and top contribution to the potential, the matrices $r$ and $l$ in eq.~(\ref{Gdef}) are needed. Since the Lagrangian for quarks is more involved, we illustrate the process with the lepton contribution. For simplicity, from eqs.~(\ref{PsiRTeven}), (\ref{newsu5multiplets}) and (\ref{newmultiplet}) we define the new multiplets
\eq{\label{multipletredefinition}
\Psi_2=\mathcal{A}\widetilde{\Psi}_2,\quad \Psi_1=\mathcal{A}\widetilde{\Psi}_1, \quad \Psi_R=\mathcal{A}\widetilde{\Psi}_R,\quad \widehat{\Psi}_R=\mathcal{B}\widetilde{\widehat{\Psi}}_R,
}
where
\eq{
\mathcal{A}=\left(\begin{array}{ccc}
-i\sigma^{2}\\
 & i\\
 &  & -i\sigma^{2}
\end{array}\right),\quad \mathcal{B}=\left(\begin{array}{ccc}
-i\sigma^{2}\\
 & i\\
 &  & 0_{2\times 2}
\end{array}\right)
}
to extract the matrices from the multiplets. Splitting the fields as in eq.~(\ref{split}) we get 
\eq{
\Lag^{(2)}_{Y_H,\widehat{Y}_H}\supset - \nn\\
\left(\begin{array}{cccc}
\overline{\psi}_{2} & \overline{\psi}_{1} & \bullet & \bullet \end{array}\right)&\left(\begin{array}{cccc}
0 & 0 & \kappa f \mathcal{A}^{\dagger}\xi \mathcal{A} & \widehat{\kappa}f \mathcal{A}^{\dagger}\widehat{\xi}\mathcal{B}\\
0 & 0 & \kappa f \mathcal{A}^{\dagger}\Sigma_0\xi^{\dag}\mathcal{A} & -\widehat{\kappa} f \mathcal{A}^{\dagger}\Sigma_0\hat{\xi}^{\dag}\mathcal{B}\\
0 & 0 & 0 & 0\\
0 & 0 & 0 & 0
\end{array}\right) \left(\begin{array}{c}
\bullet\\
\bullet\\
\psi_{R}\\
\widehat{\psi}_{R}
\end{array}\right)\nn\\&+\hc
}
where $\psi_1$, $\psi_2$, $\psi_R$ and $\widehat{\psi}_R$ are the quantum fluctuations of $\widetilde{\Psi}_1$, $\widetilde{\Psi}_2$, $\widetilde{\Psi}_R$ and $\widetilde{\widehat{\Psi}}_R$ respectively and the bullet means that the corresponding field is not present in the theory. All the flavor dependence is encoded in the couplings $\kappa$ and $\widehat{\kappa}$. Comparing with eqs.~(\ref{actionparametrization}) and (\ref{Gdef}), the form of the matrices $r$ and $l$ for leptons is
\eq{
r_l=l_l^{\dagger}=\left(\begin{array}{cccc}
0 & 0 & \kappa f \mathcal{A}^{\dagger}\xi \mathcal{A} & \widehat{\kappa}f \mathcal{A}^{\dagger}\hat{\xi}\mathcal{B}\\
0 & 0 & \kappa f \mathcal{A}^{\dagger}\Sigma_0\xi^{\dag}\mathcal{A} & -\widehat{\kappa} f \mathcal{A}^{\dagger}\Sigma_0\hat{\xi}^{\dag}\mathcal{B}\\
0 & 0 & 0 & 0\\
0 & 0 & 0 & 0
\end{array}\right).
}
To evaluate the quadratically divergent contribution to the potential arising from this sector, we need the product $ l r$
\eq{
\small
l_l r_l = \left(\begin{array}{cccc}
0 & 0 & 0 & 0\\
0 & 0 & 0 & 0\\
0 & 0 & 2\kappa^{\dag}\kappa f^2 1_{5\times 5} & \kappa^{\dag}\widehat{\kappa}f^2 \mathcal{A}^{\dag}\left(\xi^{\dag}\hat{\xi}-\xi\hat{\xi}^{\dag}\right)\mathcal{B}\\
0 & 0 & \widehat{\kappa}^{\dag}\kappa f^2 \mathcal{B}^{\dag}\left(\hat{\xi}^{\dag}\xi-\hat{\xi}\xi^{\dag}\right)\mathcal{A} & 2\widehat{\kappa}^{\dag}\widehat{\kappa} f^2 1_{3\times 3}
\end{array}\right).
}
As in the gauge boson case, the matrix $lr$ is block diagonal and its trace is the sum of the trace of each of its diagonal entries,
\eq{
\textrm{tr}\left(l_l r_l\right)= 10\textrm{tr}\left(\kappa^{\dag}\kappa f^2\right) + 6\textrm{tr}\left(\widehat{\kappa}^{\dag}\widehat{\kappa} f^2\right),
}
which is independent of the scalar fields. In this model, all the scalars are protected from quadratic divergences coming from the new sector. Analogously, for the logarithmic divergences we have to evaluate
\eq{
\textrm{tr}\left(l_l r_l l_l r_l\right)&= 2\textrm{tr}\left(\kappa \kappa^{\dag}\widehat{\kappa}\widehat{\kappa}^{\dag}f^4\right)\nn\\ &\quad\times\textrm{tr}\left[\left(\hat{\xi}^{\dag}\xi-\hat{\xi}\xi^{\dag}\right)\left(\xi^{\dag}\hat{\xi}-\xi\hat{\xi}^{\dag}\right)\mathcal{B} \mathcal{B}^{\dag}\right]\nn\\
&\supset -2 \textrm{tr}\left(\kappa \kappa^{\dag}\widehat{\kappa}\widehat{\kappa}^{\dag}f^4\right)\textrm{tr}\left[\left(\widehat{\Sigma}^{\dag}\Sigma+\Sigma^{\dag}\widehat{\Sigma}\right)\mathcal{B} \mathcal{B}^{\dag}\right]\nn\\
&= -2 \textrm{tr}\left(\kappa \kappa^{\dag}\widehat{\kappa}\widehat{\kappa}^{\dag}f^4\right)
\sum_{a=1}^3\sum_{b=1}^5
\left[\widehat{\Sigma}_{a b}^{\dag}\Sigma_{b a}+\Sigma^{\dag}_{a b}\widehat{\Sigma}_{b a}\right],
}
using that $\hat{\xi}$ commutes with $\mathcal{B} \mathcal{B}^{\dag}$ and $\Sigma_0^2=1_{5\times 5}$. Then, the logarithmically divergent contribution to the potential is
\eq{
\mathcal{V}_{L=1,l}^{\log\Lambda^2}&=-\frac{1}{8\pi^2}\log\Lambda^2\textrm{tr}\left(\kappa \kappa^{\dag}\widehat{\kappa}\widehat{\kappa}^{\dag}f^4\right) \nn\\ &\quad\times
\sum_{a=1}^3\sum_{b=1}^5
\left[\widehat{\Sigma}_{a b}^{\dag}\Sigma_{b a}+\Sigma^{\dag}_{a b}\widehat{\Sigma}_{b a}\right].
}
This expression contains leading order contributions to the $\mu^2$ parameter of the Higgs potential and to the masses of $\widehat{\omega}$ and $\widehat{\eta}$
\eq{\label{omegamass2}
\mathcal{V}^{\log\Lambda^2}_{L=1,l}&\supset \frac{f^2}{8\pi^2}\log\Lambda^2 \textrm{tr}(\kappa \kappa^{\dag}\widehat{\kappa}\widehat{\kappa}^{\dag}) \nn\\ &\quad\times\left(6 H^{\dag}H+\frac{36 \widehat{\eta }^2}{5}+8 \textrm{tr}(\widehat{\omega}^{\dagger}\widehat{\omega})\right),
}
in agreement with the diagrammatic calculation for the Higgs part in eq.~(\ref{higgs mass}).

The last contribution comes from the heavy quarks in eqs.~(\ref{Lagkappaq}), (\ref{Lagkappahatq}) and the top sector in eq.~(\ref{toplag}). In this last equation notice that due to the presence of the three dimensional Levi-Civita tensor $\epsilon_{i,j,k}$ with $\left\{i,j,k\right\}=1,2,3$, only the three upper components of $Q_{1}$ and $\Sigma_0\Omega Q_2$ are relevant. Then, comparing eqs.~(\ref{quarkmultiplets}) and (\ref{quarktopmultiplets}) we have that for $i=1,2$ one could substitute $\left(Q_{1,2}\right)_i$ by $\left(\Psi^q_{1,2}\right)_i$ and for $i=3$ we would have $i\dis T'_{1L}$ and $i\dis T'_{2L}$. Then, similarly as we did for leptons, we perform the following substitutions
\eq{
\Psi_{1}^q = \mathcal{A}\widetilde{\Psi}^q_{1}, \quad \Psi_{2}^q = \mathcal{A}\widetilde{\Psi}^q_{2},\nn\\
Q_{1}\rightarrow \mathcal{C}\widetilde{\Psi}^q_{1},\quad Q_{2}\rightarrow \mathcal{C}\widetilde{\Psi}^q_{2}
}
where 
\eq{
\mathcal{C}=\left(\begin{array}{ccc}
-i\sigma^{2} & 0 & 0\\
0 & 0 & 0\\
0 & 0 & -i\sigma^{2}
\end{array}\right)
}
since only the mentioned components of those multiplets are relevant. The zero in the middle of $\mathcal{C}$ is to take into account that the multiplets $Q_{1,2}$ and $\Psi^q_{1,2}$ defer in the field in its center. Collecting the left-handed and right-handed quantum fields in vectors
\eq{
v_L^T &= \left(\psi^{q T}_2,\psi^{q T}_1,\bullet,\bullet,\bullet,t'^T_{2L},t'^T_{1L}\right),\nn\\
v^T_R &=\left(\bullet, \bullet,\psi^{q T}_R,\widehat{\psi}^{qT}_R,\mathfrak{t}_R^T,t'^T_{2R},t'^T_{1R}\right),
}
we may write
\eq{
\mathcal{L}^{(2)}_{Y_{qH},\widehat{Y}_{qH},t}=-\overline{v}_L r_q v_R +\hc
}
where 
\eq{
r_q=\left(\begin{array}{ccccccc}
0 & 0 & \left(r_{2R}\right)^{\quad\spc\alpha}_{m n \beta} & \left(r_{2\widehat{R}}\right)^{\quad\spc\alpha}_{m n \beta} & \left(r_{2t}\right)^{\quad\alpha}_{m \beta} & 0 & 0\\
0 & 0 & \left(r_{1R}\right)^{\quad\spc\alpha}_{m n \beta} & \left(r_{1\widehat{R}}\right)^{\quad\spc\alpha}_{m n \beta} & \left(r_{1t}\right)^{\quad\alpha}_{m \beta} & 0 & 0\\
0 & 0 & 0 & 0 & 0 & 0 & 0\\
0 & 0 & 0 & 0 & 0 & 0 & 0\\
0 & 0 & 0 & 0 & 0 & 0 & 0\\
0 & 0 & 0 & 0 & \left(r_{2't}\right)^{\spc\alpha}_{\beta} & \left(r_{2'2'}\right)^{\spc\alpha}_{\beta} & 0\\
0 & 0 & 0 & 0 & \left(r_{1't}\right)^{\spc\alpha}_{\beta} & 0 & \left(r_{1'1'}\right)^{\spc\alpha}_{\beta}
\end{array}\right)
}
where the Greek indices are the SU(3) color indices and as before $l_q=r_q^{\dagger}$. As announced, this sector is more involved and it manifests in the size of the matrix $r_q$. This is because apart of the introduction of the right-handed singlets $T'_{1,2R}$ the field in the center of the \su5 multiplets $\Psi^q_{1,2}$ $\left(\chi^q_{1,2L}\right)$ is not the same as that inside of $Q_{1,2}$ $\left(T'_{1,2L}\right)$. The components of the matrix $r_q$ are defined as 
\eq{
\left(r_{2R}\right)^{\quad\spc\alpha}_{m n \beta}&=\kappa_q f \left(\mathcal{A}^{\dagger}\xi \mathcal{A}\right)_{m n}\delta_{\beta}^{\alpha},\\
\left(r_{2\widehat{R}}\right)^{\quad\spc\alpha}_{m n \beta}&=\widehat{\kappa}_q f \left(\mathcal{A}^{\dagger}\xi \mathcal{B}\right)_{m n}\delta_{\beta}^{\alpha},\\
\left(r_{2t}\right)^{\quad\alpha}_{m \beta}&= \frac{i}{4}\lambda_{1}f\left(\mathcal{C}^{\dagger}\Sigma_0\Omega\right)_{mi}\epsilon_{ijk}\epsilon_{xy}\tilde{\Sigma}_{jx}\tilde{\Sigma}_{ky}\delta_{\beta}^{\alpha},\\
\left(r_{1R}\right)^{\quad\spc\alpha}_{m n \beta}&=\kappa_q f \left(\mathcal{A}^{\dagger}\Sigma_0\xi^{\dagger} \mathcal{A}\right)_{m n}\delta_{\beta}^{\alpha},\\
\left(r_{1\widehat{R}}\right)^{\quad\spc\alpha}_{m n \beta}&=-\widehat{\kappa}_q f \left(\mathcal{A}^{\dagger}\Sigma_0\xi^{\dagger} \mathcal{B}\right)_{m n}\delta_{\beta}^{\alpha},\\
\left(r_{1t}\right)^{\quad\alpha}_{m \beta}&= \frac{i}{4}\lambda_{1}f\mathcal{C}_{mi}^{\dagger}\epsilon_{ijk}\epsilon_{xy}\Sigma_{jx}\Sigma_{ky}\delta_{\beta}^{\alpha},\\
\left(r_{2't}\right)^{\spc\alpha}_{\beta}&= \frac{1}{4}\lambda_{1}f\epsilon_{3jk}\epsilon_{xy}\tilde{\Sigma}_{jx}\tilde{\Sigma}_{ky}\delta_{\beta}^{\alpha},\\
\left(r_{1't}\right)^{\spc\alpha}_{\beta}&= \frac{1}{4}\lambda_{1}f\epsilon_{3jk}\epsilon_{xy}\Sigma_{jx}\Sigma_{ky}\delta_{\beta}^{\alpha},\\
\left(r_{2'2'}\right)^{\spc\alpha}_{\beta}&= \frac{\lambda_2}{\sqrt{2}}f\delta_{\beta}^{\alpha},\\
\left(r_{1'1'}\right)^{\spc\alpha}_{\beta}&= \frac{\lambda_2}{\sqrt{2}}f\delta_{\beta}^{\alpha}.
}
Proceeding similarly as for leptons, the quadratically divergent part of the potential due to quarks coming from the product $ l r$ reads
\eq{
\mathcal{V}^{\Lambda^2}_{L=1,q}&=-\frac{3\Lambda^2}{128\pi^2}\lambda_1^2  f^2\epsilon_{ijk}\epsilon_{inp}\epsilon_{xy}\epsilon_{qr}  \nn\\ &\quad\times\left(\Sigma_{jx}\Sigma_{ky}\Sigma^{\dagger}_{nq}\Sigma^{\dagger}_{pr}+\widetilde{\Sigma}_{jx}\widetilde{\Sigma}_{ky}\widetilde{\Sigma}^{\dagger}_{nq}\widetilde{\Sigma}^{\dagger}_{pr}\right).
}
where the factor 3 comes from $\delta^{\alpha}_{\alpha}=N_C=3$. This term contains a contribution to the triplet $\Phi$ mass and to the quartic Higgs coupling,
\eq{\label{phimass2}
\mathcal{V}^{\Lambda^2}_{L=1,q}\supset \frac{3\lambda_1^2}{4\pi^2}\Lambda^2\textrm{tr}\left(\Phi^{\dagger}\Phi\right)+\frac{3\lambda_1^2}{16\pi^2}\frac{\Lambda^2}{f^2}\left(H^{\dagger}H\right)^2.
}
From the product $l r l r$ we get the quark contribution to the logarithmic part of the potential given by
\eq{
\mathcal{V}^{\log\Lambda^2}_{L=1,q}&=\frac{3}{16\pi^2}\log\Lambda^2\Big[\frac{1}{16}\lambda_1^2\lambda_2^2f^4\epsilon_{3jk}\epsilon_{3mn}\epsilon_{xy}\epsilon_{pq}
\nn\\ &\qquad\times\left(\Sigma_{jx}\Sigma_{ky}\Sigma^{\dagger}_{mp}\Sigma^{\dagger}_{nq}+\widetilde{\Sigma}_{jx}\widetilde{\Sigma}_{ky}\widetilde{\Sigma}^{\dagger}_{mp}\tilde{\Sigma}^{\dagger}_{nq}\right)\nn\\
&+\frac{1}{16^2}\lambda_1^4f^4 \Big(\epsilon_{ijk}\epsilon_{imn}\epsilon_{xy}\epsilon_{pq}\Sigma^{\dagger}_{jk}\Sigma^{\dagger}_{ky}\Sigma_{mp}\Sigma_{nq} \nn\\ & \qquad+\epsilon_{ijk}\epsilon_{imn}\epsilon_{xy}\epsilon_{pq}\widetilde{\Sigma}^{\dagger}_{jk}\widetilde{\Sigma}^{\dagger}_{ky}\widetilde{\Sigma}_{mp}\widetilde{\Sigma}_{nq}\Big)^2\nn\\
&+\frac{1}{8}\left(\kappa_q\kappa^{\dagger}_q\right)_{33}\lambda_1^2 f^4\epsilon_{ijk}\epsilon_{ij'k'}\epsilon_{xy}\epsilon_{x'y'}\Big(\tilde{\Sigma}_{jx}\tilde{\Sigma}_{ky}\tilde{\Sigma}^{\dagger}_{j'x'}\tilde{\Sigma}^{\dagger}_{k'y'}\nn\\
&\qquad+\Sigma_{jx}\Sigma_{ky}\Sigma^{\dagger}_{j'x'}\Sigma^{\dagger}_{k'y'}\Big)\nn\\
&-2\textrm{tr}\left(\kappa_q \kappa^{\dag}_q\widehat{\kappa}_q\widehat{\kappa}^{\dag}_qf^4\right)
\sum_{a=1}^3\sum_{b=1}^5
\left(\widehat{\Sigma}_{a b}^{\dag}\Sigma_{b a}+\Sigma^{\dag}_{a b}\widehat{\Sigma}_{b a}\right)\Big],
}
where the factor $3$ comes from the number of colors. This term contains a \textit{negative} contribution to the $\mu^2$ parameter of the Higgs potential from the first term in brackets and a contribution similar to leptons up to factor 3 coming from the last term
\eq{\label{toptohiggs}
\mathcal{V}^{\log\Lambda^2}_{L=1,q}\supset & -\frac{3}{16\pi^2}\log\Lambda^2 f^2\lambda_1^2\lambda_2^2\left(H^{\dagger}H\right)\nn\\
&+\frac{3f^2}{8\pi^2}\log\Lambda^2 \textrm{tr}(\kappa_q\kappa^{\dag}_q\widehat{\kappa}_q\widehat{\kappa}^{\dag}_q) \nn\\ &\quad\times\left(6 H^{\dag}H+\frac{36 \widehat{\eta }^2}{5}+8 \textrm{tr}(\widehat{\omega}^{\dagger}\widehat{\omega})\right).
}

\subsubsection{Physical scalar masses and Higgs potential\label{BFM-physcalars}}

\begin{figure*}
\centering
\includegraphics[scale=0.39]{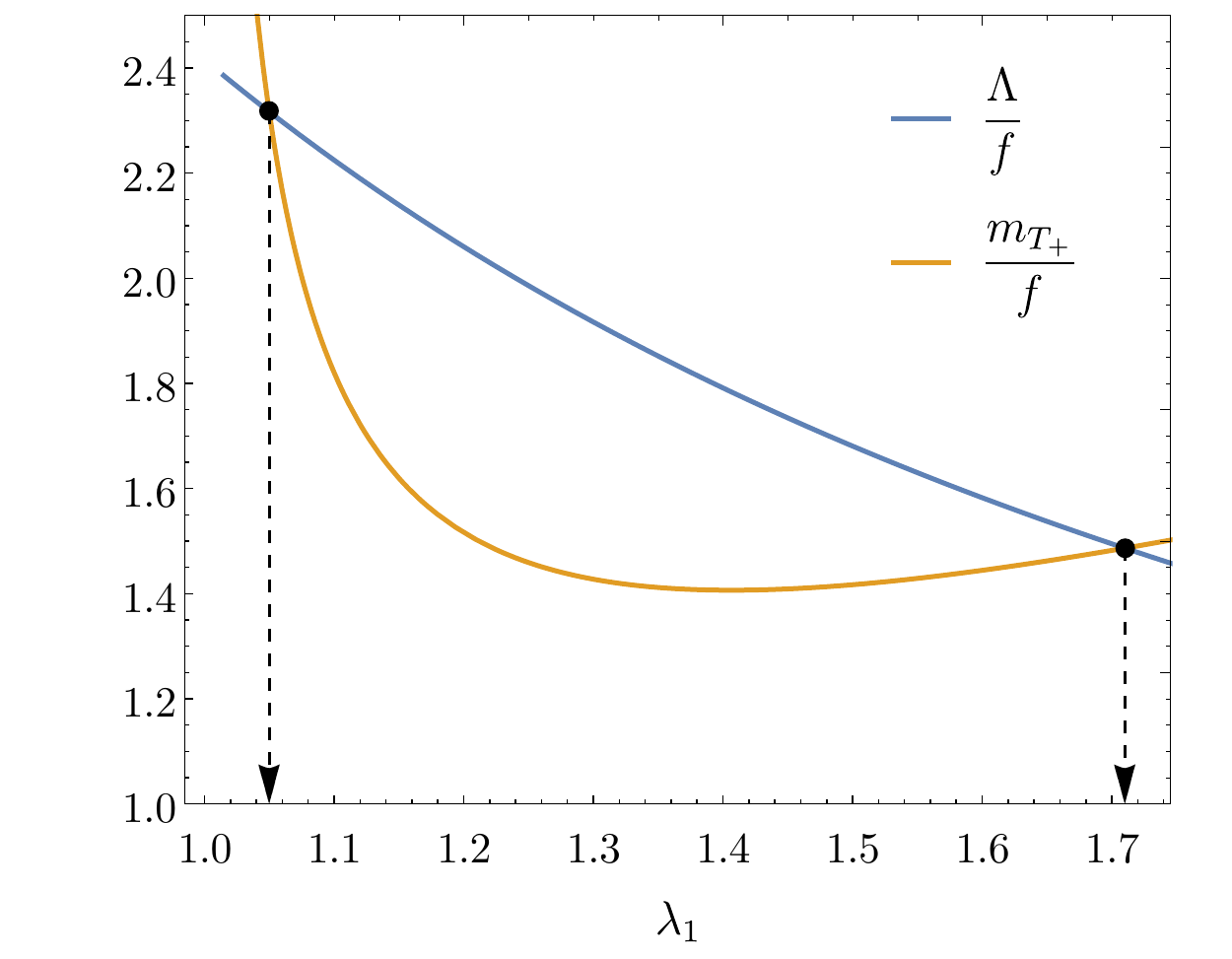}\qquad
\includegraphics[scale=0.39]{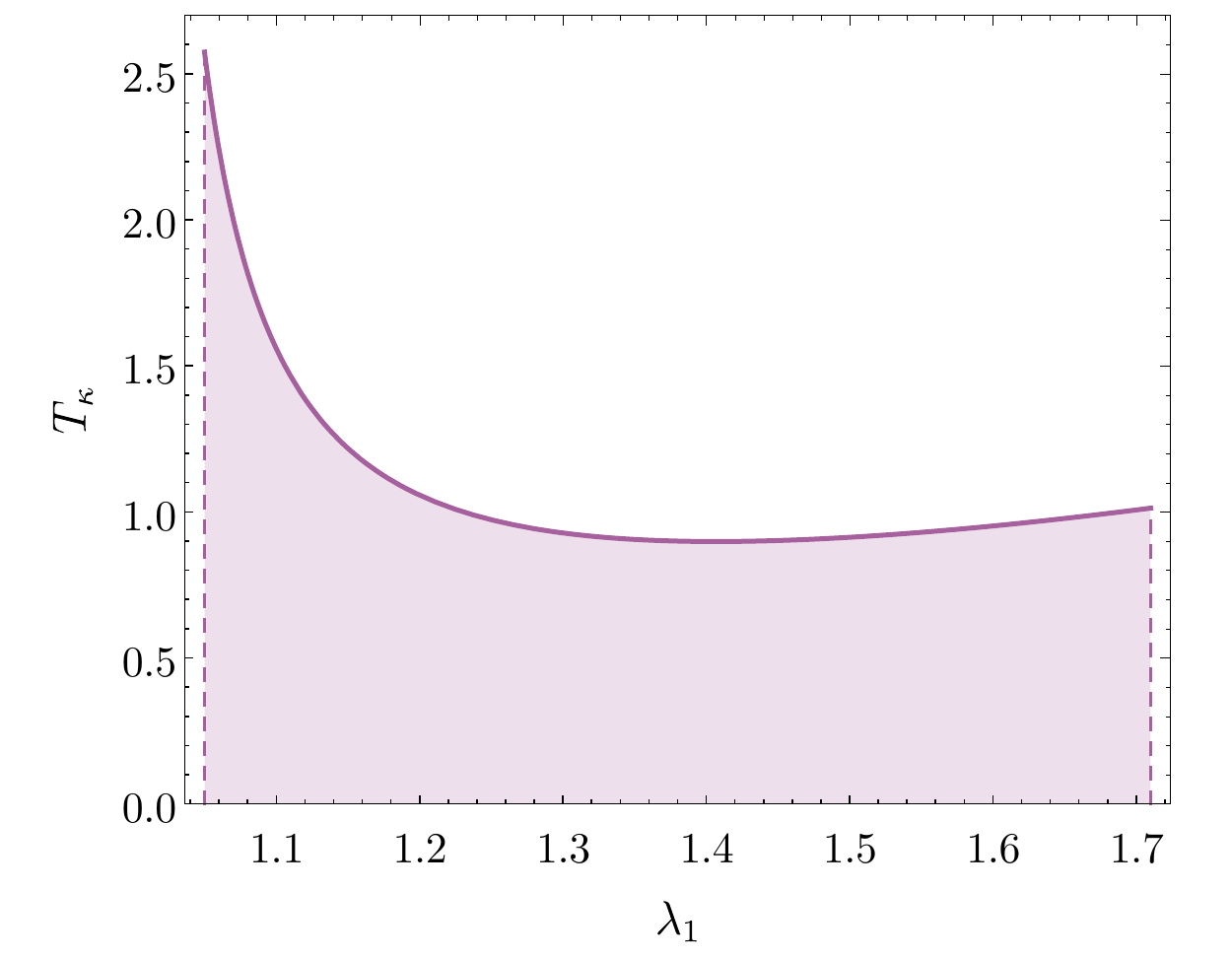}
\caption{The interval of $\lambda_1$ yielding a top quark partner mass $m_{{\rm T}_+}$ below the scale $\Lambda$ on the plot of the left-hand side determines the range of possible values of the combination of Yukawa couplings $T_\kappa\equiv\textrm{tr}(\kappa\kappa^\dagger\widehat{\kappa}\widehat{\kappa}^\dagger)
+3\textrm{tr}(\kappa_q\kappa_q^\dagger\widehat{\kappa}_q\widehat{\kappa}_q^\dagger)$ compatible with $\mu^2<0$ on the plot of the right-hand side.
\label{fig5}}
\end{figure*}

We can finally collect our results for the physical scalar masses and the Higgs potential at one loop. From eqs.~(\ref{phimass1}) and (\ref{phimass2}) we find the mass of the heaviest T-odd triplet $\Phi$
\eq{\label{phimass}
M^2_{\Phi}=\frac{\Lambda^2}{4\pi^2}\left(g^2+g'^2+3\lambda_1^2\right)
}
and the Higgs quartic coupling
\eq{
\label{lambda}
\lambda=\frac{1}{16\pi^2}\frac{\Lambda^2}{f^2}\left(g^2+g'^2+3\lambda_1^2\right).
}
The mass for the lightest T-odd triplet $\widehat{\omega}$ is given by eqs.~(\ref{omegamass1}), (\ref{omegamass2}), (\ref{toptohiggs})
\eq{
M^2_{\widehat{\omega}}=\frac{f^2}{\pi^2}\log\Lambda^2\left[g^4+T_\kappa\right],
}
with
\eq{
T_\kappa\equiv\textrm{tr}(\kappa\kappa^\dagger\widehat{\kappa}\widehat{\kappa}^\dagger)
+3\textrm{tr}(\kappa_q\kappa_q^\dagger\widehat{\kappa}_q\widehat{\kappa}_q^\dagger).
\label{eq:T}
}
The lightest T-odd scalar $\widehat{\eta}$ only receives a contribution from eqs.~(\ref{omegamass2}) and (\ref{toptohiggs}),
\eq{
M^2_{\widehat{\eta}}=\frac{f^2}{\pi^2}\log\Lambda^2\frac{9}{5}T_\kappa.
}
And the $\mu^2$ parameter of the Higgs potential follows from eqs.~(\ref{omegamass1}), (\ref{omegamass2}) and (\ref{toptohiggs}),
\eq{
\label{mu2}
\mu^2=&\frac{f^2}{16\pi^2}\log\Lambda^2\left(6g^4+\frac{2}{5}g'^4-3\lambda_1^2\lambda_2^2
+12T_\kappa\right).
}
To obtain the physical mass of the Higgs boson we have to minimize the potential of eq.~(\ref{Hpot}). If the contribution of the top sector dominates over the heavy Yukawa and gauge interactions then $\mu^2<0$ and the EWSB is triggered,
\eq{
\frac{\partial \mathcal{V}_{\textrm{Higgs}}}{\partial H^{\dagger}}=0
\quad\Rightarrow\quad \mu^2H+2\lambda\left(H^{\dagger}H\right)H=0
}
when the neutral component of the Higgs doublet gets a \vev
\eq{
\left\langle H \right\rangle = \frac{1}{\sqrt{2}}\left(\begin{array}{c}
0\\v
\end{array}\right), \quad
v = \sqrt{\frac{-\mu^2}{\lambda}}.
}
In our case, this expression gives
\eq{\label{vev}
v^2=\frac{f^4}{\Lambda^2}\log\Lambda^2
\frac{3\lambda_1^2\lambda_2^2-6g^4-\frac{2}{5}g'^4-12T_\kappa}{g^2+g'^2+3\lambda_1^2}.
}
Then, from $M^2_h= -2\mu^2= 2\lambda v^2$, and eqs.~(\ref{lambda}) and (\ref{vev}) we have
\eq{
\label{hmass}
M^2_h&=
\frac{f^2}{8\pi^2}\log\Lambda^2\left(3\lambda_1^2\lambda_2^2-6g^4-\frac{2}{5}g'^4-12T_\kappa\right),
}
whose value is $M_h\simeq125$~GeV.
Comparing previous expressions, we find the same relation between the masses of the Higgs and the heaviest triplet as in the original LHT model,
\eq{
M^2_{\Phi}=2\frac{f^2}{v^2}M^2_h,
\label{mPhi}
}
and the following one for the new physical scalar masses,
\eq{
M^2_{\widehat{\omega}}&=8M^2_h\,
\frac{g^4+T_\kappa}{3\lambda_1^2\lambda_2^2-6g^4-\frac{2}{5}g'^4
-12T_\kappa},
\label{mw}
\\
M^2_{\widehat{\eta}}&=\frac{72}{5}M^2_h\,
\frac{T_\kappa}{3\lambda_1^2\lambda_2^2-6g^4-\frac{2}{5}g'^4
-12T_\kappa}.
\label{meta}
}
Let us take $g^2\simeq 0.40$, $g'^2\simeq 0.12$, $v\simeq$~246 GeV, $m_t\simeq 173$~GeV and the relation $\lambda_1^{-2}+\lambda_2^{-2}\approx (v/(\sqrt{2}m_t))^2$ \cite{Han:2003wu} \footnote{Notice the extra factor $\sqrt{2}$ multiplying the top and top partner masses due to the different definitions of the top Yukawa couplings $\lambda_1$ and $\lambda_2$ in eq.~(\ref{toplag}) with respect to \cite{Han:2003wu}.}. The combination of Yukawa couplings $T_\kappa$ in eq.~(\ref{eq:T}) has an upper bound from Eq.~(\ref{mu2}) depending on $\lambda_1$ or $\lambda_2$ (correlated) to ensure $\mu^2<0$, and Eq.~(\ref{lambda}) provides the value of the ratio $\Lambda/f$ as a function of $\lambda_1$ given $\lambda=M_h^2/(2v^2)=0.13$. Furthermore, the cutoff scale must be $\Lambda>f$ and also greater than any particle mass or else the model would not make sense. In particular, the top quark partner ${\rm T}_+$ of mass
$m_{{\rm T}_+}=\frac{f}{\sqrt{2}}\sqrt{\lambda_1^2+\lambda_2^2}$
is the heaviest one and has to be checked. Putting together all these constraints, that must be fulfilled also by the original LHT model, we obtain from Fig.~\ref{fig5} the admissible interval of $\lambda_1$ and hence the upper limit of $T_\kappa$ values that are allowed: $\lambda_1\in[1.05, 1.71]$ and $T_\kappa\le 2.57$ for $\lambda_1\approx 1.05$ or $T_\kappa\lesssim 0.9$ for $\lambda_1\gtrsim 1.4$. In addition, any heavy fermion must be lighter than the cutoff scale, that implies all $\sqrt{2}\kappa_i$ and $\sqrt{2}\widehat{\kappa}_i$ be smaller than $\Lambda/f$, a function of $\lambda_1$ given by Eq.~(\ref{lambda}) depicted in the left panel of Fig.~\ref{fig5}. 

As for the scalars, the resulting mass of the usual triplet (\ref{mPhi}) is $M_\Phi\approx 0.73 f$, independent of any Yukawa couplings. However the masses of the new triplet and singlet scalars (\ref{mw}), (\ref{meta}), independent of $f$, could in principle take large values but only when $T_\kappa$ is extremely close to its maximum for a given $\lambda_1$, being otherwise naturally of the order of a few hundreds of GeV within the allowed range in the right panel of Fig.~\ref{fig5}.

Therefore there is enough room in the parameter space for the validity of the model below the cutoff scale.

\section{Conclusions and outlook}
\label{Conclusions}

We have examined carefully the Littlest Higgs model with T-parity, which is an interesting effective field theory that addresses the hierarchy problem justifying the lightness of the Higgs boson mass with respect to the cut-off by assuming it is a pseudo-Goldstone boson of a spontaneously broken approximate global symmetry. We have identified a couple of flaws in the fermion sector of the model, related to the mass terms of heavy fermions, that have to do with the non-trivial relation between T-parity and gauge invariance. We have probed different realizations of T-parity in the fermionic sector, in which the \su{2} singlet field in the middle of the multiplets can be even or odd, or it could be left out. We have shown that the non-linear transformation of the right-handed multiplet under the gauge group needs all the SO(5) generators, not just those associated with the SM, what forces us to rule out the T-odd realization and rely only on complete SO(5) multiplets. The reason for this is the presence of the gauge group element $\Omega$, that only commutes with the gauge generators. As a consequence, some fermions will remain massless because in particular the usual mass terms for the $\chi$ and the mirror-partner leptons are not gauge invariant since they come from the coupling of incomplet SO(5) multiplets.

To fix these issues, we have proposed that the global symmetry group $\su{5}$ is enlarged with an extra $[\su{2}\times\u{1}]^2$ factor broken spontaneously to $[\su{2}\times\u{1}]$ by the \vev\ of a new non-linear sigma field with four scalars. This allows us to introduce fermion fields that only transform in this additional non-linear representation without invoking again SO(5) multiplets that would need to be complete by gauge invariance. As a gauged subgroup we take the sum of two $[\su{2}\times\u{1}]^2$ factors, the one inside SU(5) and the extra one, hence preserving the number of gauge boson fields. This is a natural extension of the model, which was already $\su{5}\times[\u{1}]^2$, with the external abelian factors required to accommodate the hypercharges of the right-handed charged leptons. 

Once the global and gauged groups were defined we have explored two different options. In a first attempt, the left-handed components of the mirror-partner leptons and the $\chi$ were introduced in a representation that only transforms under the SM gauge group, coupled to their right-handed counterparts through both the original and new non-linear sigma fields. Then we tried with a model based on the completion of the SU(5) multiplets with new left-handed fields and the introduction of the additional right-handed components in a representation that only transforms under the SM gauge group. The first proposal, despite of being more economical in terms of fermion fields, had to be discarded because the remaining symmetry is not enough to protect the Higgs mass from quadratic divergences, as we proved with a diagrammatic calculation. However, the model with complete SU(5) multiplets is viable because it prevents all scalar fields from quadratic divergences: 
if the coupling giving masses to the extra fermion fields is switched off the Lagrangian remains SU(5) invariant. In fact the Higgs mass squared only presents an admissible logarithmic divergence proportional to $\kappa^2\widehat{\kappa}^2$, involving the product of two different couplings giving masses to the non-standard fermions hence respecting the collective symmetry breaking.

Next we have found the mass eigenfields that diagonalize the Lagrangian up to order $v^2/f^2$ as well as the fermion masses and flavor mixing matrices parametrizing the misalignment of the Yukawa couplings ($\kappa,\widehat{\kappa}$ and $\lambda$) in the flavor space of several fermion families. This version of the LHT keeps one of the original sources of lepton flavor violation \cite{blankeRareCPViolatingDecays2007,delAguila:2008zu,delAguila:2010nv} (the mixing matrix ${\sf V}$ in eq.~(\ref{mixingmatrices})), eliminates those found in \cite{delaguilaLeptonFlavorChanging2017,delaguilaInverseSeesawNeutrino2019} (now ${\sf W}={\sf Z}=1$) and introduces an additional source ($\widehat{\sf W}$) related to the new Yukawa coupling $\widehat{\kappa}$ connecting the original to the extra fermion sector.

In addition, we have considered the influence of the new fermion fields in LFV Higgs decays. Besides the contribution of the T-even right-handed singlet $(\chi_+)_R$, which is finite on its own, the contribution of the remaining fields, including the T-odd singlet, is finite. This is because they enter in the loops only through two insertions of their mixing term with the original fields of the LHT model, thus reducing the degree of divergence of the topologies involved in the process.

In the last section, we have applied the background field method to calculate the Coleman-Weinberg potential for the scalar fields generated by integrating out at one loop vector bosons and fermions, including both heavy quarks and leptons. We have calculated the one-loop contributions to the masses of the Higgs and the complex triplet of the original LHT model, as well as those of the new scalars. In our model, the Higgs mass is still not sensitive to quadratic divergences coming from the heavy leptons and heavy quark sectors. On the other hand, the relation between the Higgs mass and the complex triplet mass remains the same. Besides, the Higgs quartic coupling generated al leading order from the quadratically divergent terms of the potential does not receive contributions from the new sector. Finally, the masses of the new physical scalars are found to be proportional to just the logarithm of the high energy scale $\Lambda$. This is because they inherit part of the symmetry from the would-be Goldstone bosons to be eaten after the SSB at the scale $f$.

As a future work, we plan to extend previous phenomenological studies on lepton flavor changing processes in the context of the LHT model \cite{delaguilaLeptonFlavorChanging2017,delaguilaFullLeptonFlavor2019} ($Z$ and Higgs decays, two and three body lepton decays, $\mu\rightarrow e$ conversion in nuclei) to include the contributions from the new fermion and scalar fields. A mechanism to accommodate neutrino masses in the new LHT model, completing the work in \cite{delaguilaInverseSeesawNeutrino2019}, will also be presented elsewhere. The predictions of this model for the quark sector should be explored as well.

\section*{Acknowledgments}

We would like to thank F. del \'Aguila for his advice and a very fruitful collaboration, and T.~Hahn and J.~Santiago for ongoing work and helpful discussions. This work was supported in part by the Spanish Ministry of Science, Innovation and Universities (FPA2016-78220-C3, PID2019–107844GB-C21/AEI/10.13039/501100011033), and by Junta de Andalucía (FQM 101, SOMM17/6104/UGR, P18-FR-1962, P18-FR-5057). Funding for open access charge: Universidad de Granada / CBUA.

\bibliography{biblio}
\bibliographystyle{epj}

\end{document}